\documentclass[10pt]{article}    

\usepackage{amssymb, amsmath, amsthm, mathtools} 
\usepackage{graphicx, dsfont, enumerate, color, setspace} 
\usepackage{url}
\usepackage{caption}
\usepackage{subcaption}

\usepackage{fullpage} 

\usepackage{natbib}
\bibpunct{(}{)}{;}{a}{,}{,}

\newcommand{\bZ}{\pmb{Z}}
\newcommand{\bY}{\pmb{Y}}

\newcommand{\iv}{\mathds{1}}

\newcommand{\T}{\mathcal{T}}

\newcommand{\Var}{\mbox{\text{Var}}}

\newcommand{\Vhc}{\widehat{\mbox{Var}}_{hc2}}
\newcommand{\Vh}{\widehat{\mbox{Var}}}

\newcommand{\obs}{\text{obs}}
\newcommand{\ols}{\text{ols}}

\newtheorem{theorem}{Theorem}[section]

\newtheorem{assumption}{Assumption}
\newtheorem{definition}{Definition}
\newtheorem{proposition}{Proposition}

\title{\bf Analyzing two-stage experiments \\in the presence of interference\thanks{Email: \texttt{afeller@berkeley.edu}. We thank Peter Aronow, Peng Ding, Winston Lin, Joel Middleton, Caleb Miles, Luke Miratrix, James Pustejovsky, Todd Rogers, Shruthi Subramanyam, John Ternovksi, and Elizabeth Tipton for helpful comments and discussion. We also thank the excellent research partners at the School District of Philadelphia, especially Adrienne Reitano and Tonya Wolford.}}
\author{Guillaume Basse\\Harvard \and Avi Feller\\UC Berkeley}
\date{\today}

\begin{document}

\maketitle


\begin{abstract}
Two-stage randomization is a powerful design for estimating treatment effects in the presence of interference; that is, when one individual's treatment assignment affects another individual's outcomes. Our motivating example is a two-stage randomized trial evaluating an intervention to reduce student absenteeism in the School District of Philadelphia. In that experiment, households with multiple students were first assigned to treatment or control; then, in treated households, one student was randomly assigned to treatment. Using this example, we highlight key considerations for analyzing two-stage experiments in practice. Our first contribution is to address additional complexities that arise when household sizes vary; in this case, researchers must decide between assigning equal weight to households or equal weight to individuals. We propose unbiased estimators for a broad class of individual- and household-weighted estimands, with corresponding theoretical and estimated variances. Our second contribution is to connect two common approaches for analyzing two-stage designs: linear regression and randomization inference. We show that, with suitably chosen standard errors, these two approaches yield identical point and variance estimates, which is somewhat surprising given the complex randomization scheme. Finally, we explore options for incorporating covariates to improve precision. We confirm our analytic results via simulation studies and apply these methods to the attendance study, finding substantively meaningful spillover effects.

\end{abstract}

\doublespacing


\section{Introduction}
\label{section:intro}

A common assumption in causal inference is ``no interference'' between units: one individual's outcomes are unaffected by another individual's treatment assignment~\citep{cox1958planning, rubin1980comment}. However, this assumption does not hold in settings ranging from infectious diseases to education to labor markets~\citep[for a recent review, see,][]{Halloran:2016dc}. In many of these cases, such interference is of direct substantive interest.

Two-stage randomization is a powerful design for estimating causal effects involving interference. In the setting we consider, first whole clusters (e.g., households, schools, or graph partitions) are assigned to treatment or control. Second, units within each treated cluster are randomly assigned to treatment or control, as if each treated cluster were a separate, individually-randomized experiment. This design allows researchers to assess spillover effects either by comparing untreated units in treated clusters with pure control units in control clusters or by comparing units across clusters with different proportions assigned to treatment~\citep{hudgens2012toward}.

Our motivating example is a large randomized evaluation of an intervention targeting student absenteeism among elementary and high school students in the School District of Philadelphia~\citep{Rogers_Feller_SDP}. In the original study, parents of at-risk students were randomly assigned to a direct mail intervention with tailored information about their students' attendance over the course of the year. In treated households with multiple eligible students, one student was selected at random to be the subject of the mailings, following a two-stage randomization. Substantively, this is a rare opportunity to study intra-household dynamics around student behavior. Methodologically, this presents a rich test case for understanding how to analyze two-stage experiments in practice. 

There has been substantial interest in two-stage randomization in recent years, with prominent examples in economics~\citep{crepon2013labor}, education~\citep{somers2010enhanced}, political science~\citep{sinclair2012detecting}, and public health~\citep{hudgens2012toward}, as well as closely related variants in the context of large-scale social networks~\citep{ugander2013graph}. Such designs have become especially common in development economics~\citep{angelucci2016programme}. There is also a small but growing methodological literature on analyzing two-stage experiments, including~\citet{hudgens2012toward},~\citet{liu2014large}, and~\citet{Rigdon:2015cv} in statistics; and~\citet{sinclair2012detecting} and~\citet{baird2014designing} in the social sciences.

We build on this literature by addressing three practical issues that arise in analyzing the attendance study. First, school districts are typically interested in the intervention's impact on students rather than on households; that is, districts give equal weight to each individual rather than equal weight to each household. Similarly, public health researchers administering treatment to villages of different sizes might be interested in the impact on the overall population rather than on village-level averages, especially if the treatment is more effective in larger villages. With the exception of~\citet{sinclair2012detecting}, however, existing approaches focus either on equal weights for households~\citep[e.g.,][]{hudgens2012toward} or side-step the issue by assuming households are of equal size~\citep[e.g.,][]{baird2014designing}. We propose unbiased estimators for a broad class of individual- and household-weighted estimands, with corresponding theoretical and estimated variances. We also derive the bias of a simple difference in means for estimating individual-weighted estimands. Since researchers typically estimate these two estimands with different precision~\citep[see, e.g.,][]{Athey:2016wn}, we recommend that researchers report both in practice.

Second, we connect two common approaches for analyzing two-stage designs: linear regression, which is more common in the social sciences, and randomization inference, which is more common in epidemiology and public health. We show that, with suitably chosen standard errors, regression and randomization inference yield identical point and variance estimates. These results hold for a broad class of weighted estimands. 
 We believe this equivalence will be important in practice, since the vast majority of applied papers in this area take a ``regression first'' approach to analysis that can obfuscate key inferential issues.

Lastly, we explore options for incorporating covariates to improve precision, with a focus on post-stratification and model-assisted estimation. We then confirm our analytic results via simulation studies and apply these methods to the attendance study. Overall, we find strong evidence of a spillover effect that is (depending on the scale of the outcome) roughly 60 to 80 percent as large as the primary effect. This holds across different estimands as well as with and without covariate adjustment. Accounting for spillovers therefore dramatically improves the cost effectiveness of the intervention, from around \$5 per additional day to around \$3 per additional day.

This paper proceeds as follows.
Section~\ref{section:setup} defines the two-stage randomization, sets up the notation, and discusses the relevant assumptions.
Section~\ref{section:estimands} defines the estimands of interest both for constant and varying household sizes.
Sections~\ref{section:estimation} and \ref{section:variance} deal with unbiased estimators and their variance.
Section~\ref{sect:connection-with-regression} demonstrates how we can use regression with appropriate standard errors to obtain the randomization-based estimators.
Section~\ref{section:covariate-adjustment} explores covariate adjustment.
Section~\ref{section:simulations} reports the results of extensive simulation studies.
Section~\ref{section:data} analyzes the student attendance experiment.
Section~\ref{section:discussion} concludes and offers directions for future work. 
The supplementary materials contains additional technical material and all proofs.

\subsection{Motivating example: Student absenteeism}
More than 10 percent of public school students in the United States---over five million students---are chronically absent each year, defined as missing 18 or more days of the roughly 180-day school year~\citep{crdc2016}. \citet{Rogers_Feller_SDP} recently conducted the first randomized evaluation of an intervention aimed at reducing student absenteeism for this population. This intervention delivered targeted information to parents of at-risk students in the School District of Philadelphia via five pieces of direct mail over the 2014--2015 School Year. The mailing clearly stated the student's number of absences that year (``Your student has been absent 16 days this school year''), included a simple bar chart showing the same information graphically, and gave additional text on the importance of attending school.
~\citet{Rogers_Feller_SDP} find that the treatment reduces chronic absenteeism by over 10 percent relative to control.  The approach is extremely cost-effective, costing around \$5 per additional day of student attendance---more than an order of magnitude more cost-effective than the current best-practice intervention.

A key practical challenge in implementing the original study was that the mailings were designed to provide information about a single student. Students were eligible to be the target of the intervention if they met certain pre-specified criteria, including type of school, home language, and no perfect attendance in the previous year. In households with multiple eligible students, one student was randomly selected to be the focal student. (The study excludes other members of the household, such as non-eligible siblings.)~\citet{Rogers_Feller_SDP} addressed possible spillover only briefly in the original study, largely because the focus was on the primary effect of the intervention and because households with multiple students were around 15 percent of all households in the sample.

In this paper, we consider a subset of $N = 3,876$ households with between $n_i = 2$ and $n_i = 7$ eligible students in each household and $n^+ = 8,654$ total students. Table~\ref{tbl:num_students_hh} shows the distribution of household size. The vast majority of these households (82 percent) have only two students; only one percent (35 households) have five or more students. Our goal is to estimate the primary and spillover effects on attendance for this finite sample. We are also interested in the extent to which these estimates differ depending on whether we give equal weight to each household or to each individual. The original study estimated these effects with a simple difference-in-means estimator, which could be biased in practice; see Section~\ref{section:sd-bias}.

\begin{table}[btp]
	\centering
	\caption{Number of households by size, and proportion of treated households for each size.}
	\label{tbl:num_students_hh}
	\begin{tabular}{r ccc}
	& 2 & 3 & 4--7  \\
	\hline
	Total $N$ & 3,169 & 557 & 150 \\
	Proportion assigned to treatment & 0.66 & 0.65 & 0.65 \\
	\hline
	\end{tabular}	
\end{table}

This experimental design presents a rare opportunity to assess intra-household spillovers. There is substantial evidence across fields that such intra-household spillovers are meaningful in magnitude. For example, several voter mobilization studies have found spillover effects that are between one-third and two-thirds as large as the primary effect~\citep{nickerson2008voting, sinclair2012detecting}. We are interested in spillover in the attendance study for two key reasons. First, ignoring the spillover effect under-states the overall impact of the intervention. For example, an important metric is the cost of each additional student day; ignoring spillover artificially lowers the corresponding cost-effectiveness estimates. Second, the research team faced a practical question of whether to develop a separate intervention for households with multiple eligible students, which would be costly to implement and test. If the spillover effect is comparable in magnitude to the primary effect, such development is unnecessary. This is similar to decisions around interventions targeting infectious diseases~\citep{hudgens2012toward} and in economics~\citep{baird2014designing}.

\section{Setup and assumptions\label{section:setup}}

We now review the setup and assumptions for a two-stage experiment in the presence of interference. The discussion closely follows~\citet{hudgens2012toward}, modifying their terminology slightly to better fit our applied example and to recognize some small differences in emphasis in the social science literature. We first define potential outcomes and state the relevant assumptions, then follow with a description of two-stage randomized designs. We postpone the formal introduction of our estimands to Section~\ref{section:estimands}. For additional reviews on causal inference under interference, see, among others,~\citet{sinclair2012detecting, bowers2013reasoning, VanderWeele:2014gr, Halloran:2016dc, aronow2012estimating, Athey:2016wn}.

\subsection{Potential outcomes and relaxing SUTVA}
\label{section:assumptions}
We use the potential outcomes framework to describe the 
problem~\citep{neyman::1923, rubin1974estimating}.
Consider N households $i = 1, \ldots, N$ with $n_i$ individuals in household $i$, and where $n^+ \equiv \sum n_i$
 is the total number of individuals. To be consistent with the existing literature 
 on two-stage experiments, we use the double-index notation, such that $\cdot_{ij}$ denotes 
 the individual $j$ in household $i$. For household $i$, let
$\bZ_i = (Z_{i1}, \ldots, Z_{in_i})$ denote the assigment vector for the $n_i$ units in that household, where
$Z_{ij} = 1$ if the $j^{th}$ individual in household $i$ is assigned to treatment, and $Z_{ij} = 0$ otherwise.
Similarly, define $\bZ_{i, -j}$ as the sub-vector of $\bZ_i$ that excludes the $j^{th}$ value. Finally, aggregate all 
household-level assignments via $\bZ = \{\bZ_1, \ldots, \bZ_N\}$. Let $Y_{ij}^{\text{obs}}$ denote the observed
 outcome for individual $j$ in household $i$, which will be either binary or continuous in our motivating example. 
 In general, let $\bY_i(\bZ) = (\bY_{i1}(\bZ)$, $\ldots$, $\bY_{in_i}
 (\bZ))$ be the vector of potential outcomes for household $i$, and $\bY(\bZ) = \{ \bY_1(\bZ)$, $\ldots$,
  $\bY_N(\bZ)\}$ be the list of potential outcome vectors for all households. 
  
  At this stage, practical inference
  is infeasible without imposing additional restrictions on the structure of potential outcomes. However, the standard
  Stable Unit Treatment Value Assumption~\citep[SUTVA;][]{rubin1980comment}, which implies that there is no
  interference between units, is inappropriate in our context. We instead focus on putting structure on two types of interference: between-household and within-household interference.

\subsubsection{Between-household interference}
First, we assume that there is no between-household interference, which~\citet{sobel2006randomized} refers to as \textit{partial interference}.

\begin{assumption}[No Interference Across Household] \label{asst:partial-interference}
	Interference occurs only within a household. That is, $\bY_i(\bZ) = \bY_i(\bZ_i)$.
\end{assumption}

This is effectively a ``between household SUTVA'' assumption and greatly reduces the complexity of the problem.
In the context of the attendance study, this assumption states that students in different households do not affect each others' attendance. This assumption is violated if, for instance, friends skip school together. 
Nonetheless, we view spillovers within households as far more important than spillovers between households and thus consider Assumption~\ref{asst:partial-interference} to be a useful approximation. 

\subsubsection{Within-household interference}
Even with the partial interference assumption, practical inference remains challenging. The key complication is that, without additional restrictions, the potential outcomes depend on the identity of the treated individual. To see this, consider household $i$ with three students in which only one student is assigned to treatment. Under partial interference, the oldest student, $j = 1$, has three potential outcomes, $Y_{i1}(1,0,0)$, $Y_{i1}(0,1,0)$, and $Y_{i1}(0,0,1)$, which correspond to assigning the oldest, middle, and youngest student to treatment, respectively, as well as $Y_{i1}(0,0,0)$ if none receive treatment. Thus, the oldest student actually has \textit{two} different spillover effects, $Y_{i1}(0,1,0) - Y_{i1}(0,0,0)$ and $Y_{i1}(0,0,1) - Y_{i1}(0,0,0)$, depending on which other student in the household receives the treatment.

As~\citet{hudgens2012toward} argue, this makes inference difficult, especially with respect to variance calculations~\citep[see also][]{tchetgen2012estimation}. Instead,~\citet{hudgens2012toward} propose the \textit{stratified interference} assumption, which states that the precise identity of the treated individual in the treated cluster does not matter for untreated individuals in the same cluster~\citep[see also][]{manski2013public}. 
\begin{assumption}[Stratified Interference] \label{asst:stratified-interference} 
	\begin{equation}
	Y_{ij}(\bZ_{i,-j},Z_{ij}=0)=Y_{ij}(\bZ'_{i,-j},Z_{ij}=0) \,\,\,\,\, \forall \bZ_{i,-j}, \bZ'_{i,-j} \,\,\,\, \mbox{s.t.}\,\,\, \sum_j^{n_i} Z_{ij}=\sum_j^{n_i} Z'_{ij}=1
	\end{equation}
\end{assumption}
Heuristically, this imposes additional structure on the problem by assuming that potential outcomes are only a function of the number (or, depending on context, proportion) of individuals assigned to treatment within each household.

\subsection{Assignment mechanism and observed outcomes}

\begin{figure}[btp]
		\centering
		\includegraphics[scale = 0.4]{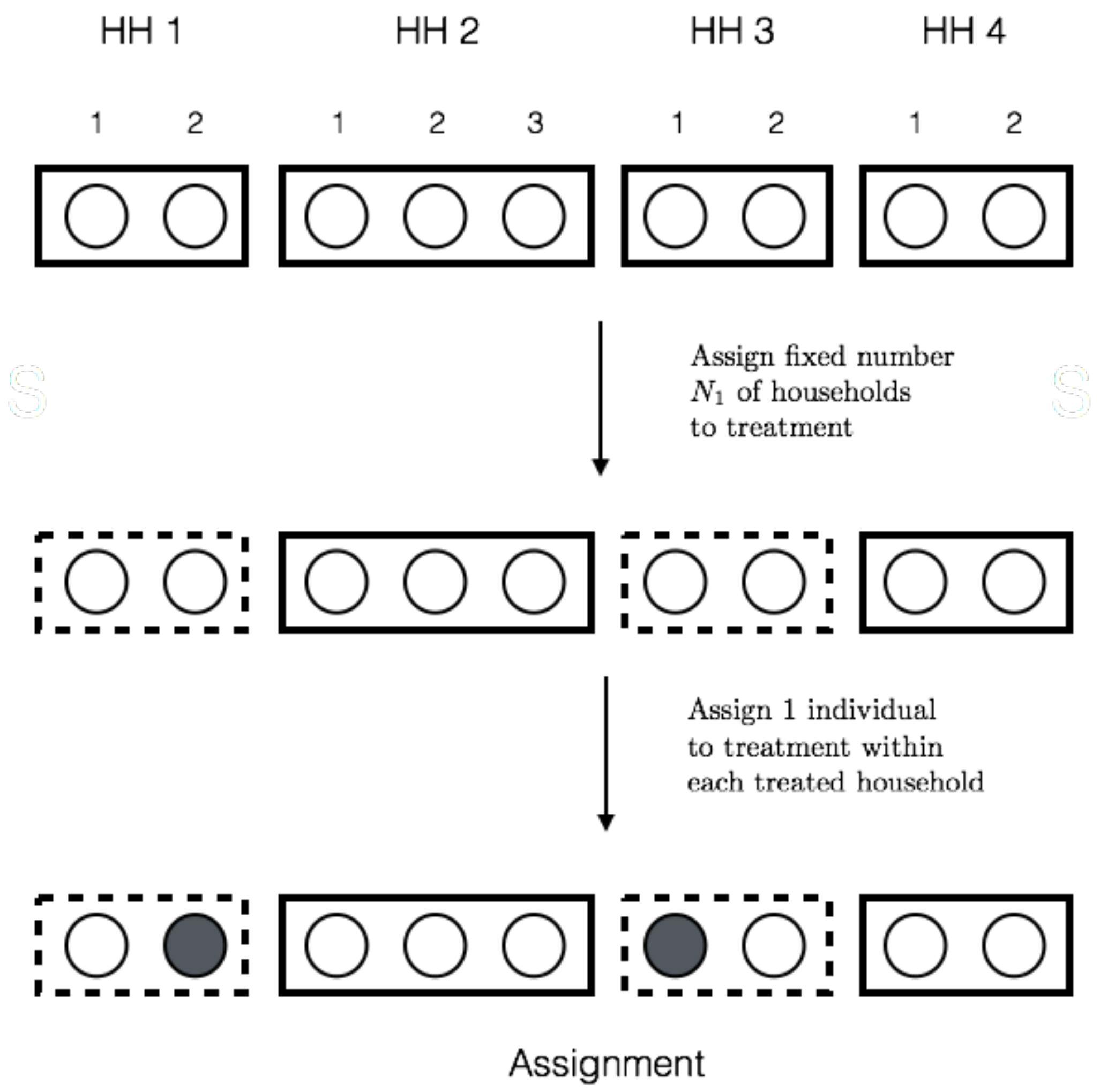}
		\caption{Schematic of the two-stage randomized design we consider. Treated
		households are represented by dashed bold rectangles. Treated individuals are 
		filled circles.}
		\label{fig:multilevel_schematic}
\end{figure}

Two-stage randomization is a special case of a multi-stage, nested randomization~\citep[see, e.g.,][]
{sobel2006randomized} that is used to assign treatment to units in a nested structure
(throughout, we will refer to individuals nested within household). Figure~\ref{fig:multilevel_schematic} 
highlights the sequential nature of the two-stage design we consider. Specifically, we consider designs in which each stage follows complete randomization; that is, randomizations in which a fixed number of units are assigned to treatment at each stage~\citep{imbens2015causal}. See also~\citet{liu2014large}, who refer to this as a permutation randomization, and~\citet{tchetgen2012estimation}, who contrast completely randomized and Bernoulli designs in the second stage. 

Formally, let $\pmb{H} = (H_1, \ldots, H_N)$ be the vector of treatment assignments at the household level, 
such that $H_i = 1$ if household $i$ is assigned to treatment and $H_i = 0$ otherwise. For the first stage of randomization, we assume that a fixed integer of households $N_1 \in \{1, \ldots, N-1\}$ are assigned to treatment, with $P(\pmb{H}) = 1/{N \choose N_1}$ for all $\pmb{H}$ such that $\sum_i^N H_i = N_1$ and $P(\pmb{H}) = 0$ otherwise. Analogously define $N_0 = N - N_1$ as the number of households assigned to control. For the second stage, individuals are assigned to treatment conditional on the realized value of the first stage of randomization. For individuals in households with $H_i = 1$, we assume that exactly one individual is assigned to treatment, with $P(\bZ_i \mid H_i = 1) = \frac{1}{n_i}$ if $\sum_j^{n_i} Z_{ij} = 1$ and $P(\bZ_i \mid H_i = 1) = 0$ otherwise. For individuals in households with $H_i = 0$, $Z_{ij} = 0$ for all $j$. Thus, all individuals assigned to treatment are in households assigned to treatment.

Given the assumptions in Section~\ref{section:assumptions} and the type of assignments allowed by the two-stage randomization mechanism we consider, the potential outcomes for individual $j$ in household $i$ 
simplify to $Y_{ij}(\bZ) = Y_{ij}(H_i = h, Z_{ij}=z)$.  There are three possible combinations: $Y_{ij}(1,1)$, $Y_{ij}(1,0)$, and $Y_{ij}(0,0)$. 
We regard these potential outcomes as fixed and define the observed outcome as a deterministic function of the treatment assignment and potential outcomes:
$$Y_{ij}^{\obs} = H_i Z_{ij} Y_{ij}(1,1) + H_i(1 - Z_{ij})Y_{ij}(1,0) + (1 - H_i)Y_{ij}(0,0),$$
where the randomness is entirely due to $\pmb{H}$ and $\bZ$. That is, unless otherwise stated, all expectations and variances are with respect to the randomization distribution; inference is fully justified by the randomization itself~\citep{fisher1960design}. Finally, we introduce the sets $\T_{hz} = \{(i,j) : H_i = h \,\, \mbox{ and } \,\,  Z_{ij} = z\}$ to denote the set of households and individuals who are assigned to $H_i = h$ and $Z_{ij} = z$.


\section{Estimands} \label{section:estimands}
We next discuss estimands of interest, closely following~\citet{hudgens2012toward}. We start with the setting in which all households or clusters are of equal size and then turn to households of varying size.

\label{section:estimands-equal}
\subsection{Constant household size}
\subsubsection{Primary and Spillover Effects}
We first assume that all households are of the same size. That is, $n_i = n$ for all households $i = 1, \ldots, N$. Next, we define average potential outcomes at the household level, $\overline{Y}_i(h,z) = \frac{1}{n}\sum_j^n Y_{ij}(h,z)$, which average across all individuals $j = 1, \ldots, n$ within each household. Since all households are the same size, we can analogously define the average potential outcomes for the sample, $\overline{Y}(h,z) = \frac{1}{N}\sum_i^N \overline{Y}_{i}(h,z)$, which is the average household-level potential outcome across all households. Estimands are contrasts between these sample average potential outcomes. Unless otherwise stated, all estimands we consider here are finite sample estimands; that is, they are defined for the units in our sample.

\begin{definition}[Estimands with equal-sized households]\label{def:equal-size-fp-effects}
	Define the average {\em primary effect} as follows:
	\begin{equation}
		\tau^P = \frac{1}{Nn} \sum_{i}^N\sum_j^n (Y_{ij}(1,1) - Y_{ij}(0,0)) = \frac{1}{N}\sum_i^N (\overline{Y}_{i}(1,1) - \overline{Y}_{i}(0,0)) = \overline{Y}(1,1) - \overline{Y}(0,0),
	\end{equation}
	and the average {\em spillover effect} as:
	\begin{equation}
		\tau^S = \frac{1}{Nn} \sum_{i}^N\sum_j^n (Y_{ij}(1,0) - Y_{ij}(0,0)) = \frac{1}{N}\sum_i^N (\overline{Y}_{i}(1,0) - \overline{Y}_{i}(0,0)) = \overline{Y}(1,0) - \overline{Y}(0,0).
	\end{equation}
\end{definition}

These estimands have various names in the literature. We take the terminology \textit{primary effect} from~\citet{toulis2013estimation}, but \citet{baird2014designing} 
use the term \textit{treatment on the uniquely treated} for an analogous quantity.~\citet{hudgens2012toward} call these estimands the \textit{total} and \textit{indirect} effects, respectively. 

\subsubsection{Other Estimands\label{section:overall-effect}}
\citet{hudgens2012toward} propose two additional estimands, the \textit{direct} and \textit{overall} effects. The direct effect is essentially the impact of the second stage, within-household randomization, which is $\tau^D = \overline{Y}(1,1) - \overline{Y}(1,0)$ with equal-sized households. Following the effect decomposition in~\citet{hudgens2012toward}, the direct effect can be defined as the difference between the primary and spillover effects. Since the spillover effect is a more natural estimand in our setting, we do not discuss the direct effect further.

The overall effect is the impact of household-level random assignment. From a policy perspective, this quantity is of obvious interest to decision makers, providing a single number for the intervention's impact. From a statistical perspective, however, the overall effect has the somewhat awkward feature that it is only defined with respect to a given assignment mechanism:
	\begin{equation}
		\tau^O = \frac{1}{Nn} \sum_{i}^N\sum_j^n \left(E\left[Y_{ij}(1,Z_{ij})\right] - Y_{ij}(0,0)\right),
	\end{equation}
	where $E\left[Y_{ij}(1,Z_{ij})\right]$ is the expected potential outcome for individual $j$ in household $i$:
	$$E\left[Y_{ij}(1,Z_{ij})\right] = P(Z_{ij} = 1 \mid H_i = 1)~Y_{ij}(1,1) + P(Z_{ij} = 0 \mid H_i = 1)~Y_{ij}(1,0).$$ 
Similar to~\citet{VanderWeele:2011fd}, we can therefore re-write the overall effect as a weighted average of the primary and spillover effects:
$$\tau^O = P(Z_{ij} = 1 \mid H_i = 1)~\tau^P + P(Z_{ij} = 0 \mid H_i = 1)~\tau^S,$$
with weights equal to the second stage treatment probability. As a result, we focus on the primary and spillover effects throughout the main text and defer corresponding results for the overall effect to the supplementary materials.

\subsection{Varying household size}
\label{sec:overall-weights}
We now generalize these results to allow for varying household size. Broadly, there are now two types of estimands, \textit{household-weighted} estimands (`HW') that assign equal weight to households, regardless of the number of individuals in each household; and \textit{individual-weighted estimands} (`IW') that assign equal weight to individuals, regardless of the distribution across households. A substantial literature on cluster-randomized trials addresses related questions; see, among others,~\citet{donner2000design, imai2009essential, schochet2013estimators,aronow_middleton_cluster}. Specifically, we generalize the equal-sized household estimands to allow for \textit{two-stage weights}.
\begin{definition}[Two-stage weighted estimands]\label{def:unequal-size-fp-effects}
	Define the average {\em primary effect} as follows:
	\begin{equation}
		\tau^P_W = \sum_{i=1}^N w_i^\ast \sum_{j=1}^{n_i} (Y_{ij}(1,1) - Y_{ij}(0,0)),
	\end{equation}
	and the average {\em spillover effect} as:
	\begin{equation}
		\tau^S_W = \sum_{i=1}^N w_i^\ast \sum_{j=1}^{n_i} (Y_{ij}(1,0) - Y_{ij}(0,0)),
	\end{equation}
	where $w_i^\ast = \frac{1}{Nn_i}$ corresponds to household-weighted estimands and $w_i^\ast = \frac{1}{n^+}$ corresponds to individual-weighted estimands.
\end{definition}

\noindent When $n_i = n$ for all $i$, both HW and IW estimands are identical to the equal-sized household estimands. When $n_i$ is not constant, the resulting household weighted estimands are:
\begin{equation*}
	\tau^P_{HW} = \frac{1}{N} \sum_{i=1}^N \frac{1}{n_i} \sum_{j=1}^{n_i} (Y_{ij}(1,1) - Y_{ij}(0,0)) \,\,\,\,\, \mbox{and}
	\,\,\,\,\, \tau^S_{HW} = \frac{1}{N} \sum_{i=1}^N \frac{1}{n_i} \sum_{j=1}^{n_i} (Y_{ij}(1,0) - Y_{ij}(0,0)).
\end{equation*}
\noindent These are the estimands in~\citet{hudgens2012toward}. The corresponding individual weighted estimands are:
\begin{equation*}
\tau^P_{IW} = \frac{1}{n^+} \sum_i^N \sum_j^{n_i} (Y_{ij}(1,1) - Y_{ij}(0,0)) \,\,\,\,\,\,\, \mbox{ and }  \,\,\,\,\,\,\,
\tau^S_{IW} = \frac{1}{n^+} \sum_i^N \sum_j^{n_i} (Y_{ij}(1,0) - Y_{ij}(0,0)),
\end{equation*}
where $n^+$ is the total number of individuals.

\section{Estimation}\label{section:estimation}
Next, we turn to estimating these quantities of interest. First, we generalize the results of~\citet{hudgens2012toward} to allow for unbiased estimation of any two-stage weighted estimand. We then discuss additional complications that arise when estimating individual-weighted effects.

\subsection{Unbiased estimation}\label{section:unbiased-estimation}
We now define two-stage weights for estimation, of which household and individual weights are special cases.

\begin{proposition}[Unbiasedness of two-stage weighted estimators]\label{prop:eq-est-unbiasedness}
	Define two-stage inverse probability weights $w_i^{(00)}, w_i^{(10)}$, and $w_i^{(11)}$ as follows:
	\begin{align*}
		w_i^{(11)} &= \frac{1}{P(H_i = 1)}\frac{1}{P(Z_{ij}=1 | H_i=1)}, \\
		w_i^{(10)} &= \frac{1}{P(H_i = 1)}\frac{1}{P(Z_{ij}=0 | H_i=1)}, \\
		w_i^{(00)} &= \frac{1}{P(H_i = 0)}.
	\end{align*}
	Consider two-stage estimand weights, $w_i^\ast$, as in Definition~\ref{def:unequal-size-fp-effects}. The weighted primary and spillover effect estimators $\widehat{\tau}^P_W$ and $\widehat{\tau}^S_W$,
		\begin{eqnarray*}
			\widehat{\tau}^P_W &=& \sum_{(i,j)\in \T_{11}} w_i^{(11)} w_i^\ast Y_{ij}^{\obs}(1,1) - \sum_{(i,j) \in \T_{00}} w_i^{(00)} w_i^\ast Y_{ij}^{\obs}(0,0),\\
			\widehat{\tau}^S_W &=& \sum_{(i,j)\in \T_{10}} w_i^{(10)} w_i^\ast Y_{ij}^{\obs}(1,0) - \sum_{(i,j) \in \T_{00}} w_i^{(00)} w_i^\ast Y_{ij}^{\obs}(0,0),
		\end{eqnarray*}
are unbiased for their corresponding estimands with respect to the randomization distribution. That is $\mathbb{E}[\widehat{\tau}^P_W] = \tau^P_W$ and $\mathbb{E}[\widehat{\tau}^S_W] = \tau^S_W.$
\end{proposition}

The proof is given in the supplementary materials and follows closely from~\citet{hudgens2012toward}. These estimators have a simple difference-in-means form either when household size is constant or with equal weight on households, as in~\citet{hudgens2012toward}:
\begin{equation}
	\widehat{\tau}^P_{HW} = \frac{1}{N_1} \sum_{i\in \T_{11}}\overline{Y}_i^{\obs}(1,1) - \frac{1}{N_0} \sum_{i \in \T_{00}} \overline{Y}_i^{\obs}(0,0) \,\,\,\,\, \mbox{and}
	\,\,\,\,\, \widehat{\tau}_{HW}^S = \frac{1}{N_1} \sum_{i\in \T_{10}}\overline{Y}_i^{\obs}(1,0) - \frac{1}{N_0} \sum_{i \in \T_{00}} \overline{Y}_i^{\obs}(0,0).
\end{equation}

\noindent The unbiased, individual weighted estimators have the form of Horvitz-Thompson estimators modified for our two-stage randomization. Thus the unbiased estimators for $\tau^P_{IW}$ and $\tau^S_{IW}$ are:
\begin{align*}
	\widehat{\tau}^P_{IW} &= \frac{1}{n^+}\frac{N}{N_1} \sum_{(i,j) \in \T_{11}}  \frac{n_i}{1} \, Y_{ij}^{\obs}(1,1) \quad\quad\;\; - \quad \frac{1}{n^+}\frac{N}{N_0} \sum_{(i,j) \in \T_{00}}  Y_{ij}^{\obs}(0,0), \\[1em]
	\widehat{\tau}^S_{IW} &=  \frac{1}{n^+}\frac{N}{N_1} \sum_{(i,j) \in \T_{10}}  \frac{n_i}{n_i - 1} \, Y_{ij}^{\obs}(1,0) \quad - \quad   \frac{1}{n^+}\frac{N}{N_0}  \sum_{(i,j) \in \T_{00}} Y_{ij}^{\obs}(0,0),
\end{align*}
with household-level assignment probabilities $N_1/N$ and $N_0/N$ and (conditional) individual-level probabilities $1/n_i$ and $(n_i-1)/n_i$. Since these are inverse probability weight estimators and since the probability of treatment assignment is a function of household size, the primary effect estimator up-weights larger households, with weights proportional to $n_i = \{1/n_i\}^{-1}$. Similarly, the spillover effect estimator down-weights larger households, with weights proportional to $n_i/(n_i - 1) = \{(n_i-1)/n\}^{-1}$. 

As is common with Horvitz-Thompson estimators, unbiased estimation typically comes at the price of additional variance. In practice, researchers can often reduce this variance by first normalizing the weights (i.e., H{\'a}jek weights), which introduces some small bias. See the supplementary materials for more details on the H{\'a}jek estimator in this context.

\subsection{Bias of the simple difference estimator} \label{section:sd-bias}
Contrast the unbiased estimator with a simple difference estimator, that is, the difference-in-means across individuals ignoring households:
\begin{align}
	\widehat{\tau}^{P}_{sd} &= \frac{1}{n_{11}^+} \sum_{(i,j) \in \T_{11}}  Y_{ij}^{\obs}(1,1) - \frac{1}{n_{00}^+} \sum_{(i,j) \in \T_{00}} Y_{ij}^{\obs}(0,0) \label{eq:primary_sd}\\
	\widehat{\tau}^{S}_{sd} &=  \frac{1}{n_{10}^+} \sum_{(i,j) \in \T_{10}} Y_{ij}^{\obs}(1,0) - \frac{1}{n_{00}^+} \sum_{(i,j) \in \T_{00}} Y_{ij}^{\obs}(0,0),\label{eq:spillover_sd}
\end{align}
where $n_{11}^+ = \sum_i \iv(H_i=1) \sum_j \iv(Z_{ij}=1)$, $n_{10}^+ = \sum_i \iv(H_i=1) \sum_j \iv(Z_{ij}=0)$, and $n_{00}^+ = \sum_i \iv(H_i=0)n_i$. These are essentially the estimators used in~\citet{Rogers_Feller_SDP}.

Despite its intuitive appeal, this estimator can be biased in practice. There are two main sources of bias. First, echoing results from~\citet{aronow_middleton_cluster}, when household sizes vary, the quantities $n_{11}^+$, $n_{10}^+$, and $n_{00}^+$ are themselves random variables. Thus, both the numerator and denominator of each group average are random; and the mean of a ratio is not, in general, equal to the ratio of means. Second, individual-level treatment probabilities vary by household size; in the design we consider here, the probability of treatment assignment conditional on being in a treated household is $\mathbb{P}\{Z_{ij} = 1 \mid H_i = 1\} = 1/n_i$. Thus, ignoring $n_i$---and, by extension, the varying treatment probability---can lead to biased estimates. 
We derive the exact form of the bias in the following proposition.

\begin{proposition}\label{prop:bias-sd-iw}
	The simple difference estimators, $\widehat{\tau}^{P}_{sd}$ and $\widehat{\tau}^{S}_{sd}$, defined in Equation~\ref{eq:primary_sd} and~\ref{eq:spillover_sd}, have the following bias for their respective estimands.

	\begin{equation}
	\text{bias}\left(\widehat{\tau}^{P}_{sd}\right) = \frac{1}{N\overline{n}} \sum_i \left( \frac{\overline{n}}{n_i} - 1\right) \sum_j Y_{ij}(1,1) +
	\frac{1}{N_0\overline{n}} \text{cov}\left( \frac{\sum_{\T_{00}} Y_{ij}(0,0)}{ n_{00}^+ } , n_{00}^+\right)
	\end{equation}
	and
	\begin{align}
	\text{bias}\left(\widehat{\tau}^{S}_{sd}\right) &= \frac{1}{N\overline{n}} \sum_i \left(\frac{ \frac{\overline{n}}{\overline{n}-1} }{\frac{n_i}{n_i-1}} - 1\right) \sum_j Y_{ij}(1,0) + 
	\bigg( \frac{1}{N_0\overline{n}} \text{cov}\left( \frac{\sum_{\T_{00}}Y_{ij}(0,0)}{n_{00}^+}, n_{00}^+\right) -\\	
	& \qquad\qquad \frac{1}{N_1(\overline{n}-1)} \text{cov}\left( \frac{\sum_{\T_{10}}Y_{ij}(1,0)}{n_{10}^+}, n_{10}^+\right) \bigg). \nonumber
	\end{align}
\end{proposition}

If household size is constant, all of these terms are zero. If the covariance between household size and potential outcomes is zero, only the first term of each equation remains. In simulations in Section~\ref{section:simulations}, we show that the overall bias can be large if household sizes vary and treatment effects also vary by household size. 

\subsection{Stratification and post-stratification by household size}
\label{section:stratified-randomization}

Finally, we consider stratification and post-stratification by household size. If household-level randomization is stratified by household size, inference for the individual-weighted estimand is immediate. In particular, let $\tau^P_{k}$ and $\tau^S_{k}$ be the stratum-specific estimands for the stratum with household size $n_i = k$,
\begin{equation*}
	\tau^P_k = \frac{1}{N^{(k)}} \sum_i^{N} \iv(n_i = k) \frac{1}{k} \sum_j^k (Y_{ij}(1,1) - Y_{ij}(0,0)) \,\,\,\,\,\, \mbox{and} \,\,\,\,\,\,
	\tau^S_k = \frac{1}{N^{(k)}} \sum_i^{N} \iv(n_i = k) \frac{1}{k} \sum_j^k (Y_{ij}(1,0) - Y_{ij}(0,0)),
\end{equation*}
where $N^{(k)}$ is the number of households of size $k$. Since household size is constant within each stratum, the corresponding household- and individual-weighted estimands are equivalent. We can therefore re-write the overall individual-weighted estimands as weighted averages of the stratum-specific effects,
\begin{equation*}
	\tau^P_{IW} = \sum_{k=2}^K \frac{n^{(k)+}}{n^+} \tau^{P}_k \,\,\,\,\,\, \mbox{ and } \,\,\,\,\,\, \tau^S_{IW} = \sum_{k=2}^K \frac{n^{(k)+}}{n^+} \tau^{S}_k,
\end{equation*}
where $n^{(k)+} = \sum_{i:~n_i = k} n_i$ and where we assume (without essential loss of generality) that household sizes range from $k=2,\ldots, K$. Plugging in $\widehat{\tau}^{P}_k$ and $\widehat{\tau}^{S}_k$ gives the corresponding unbiased estimate.

%
To modify the above results for household-weighted estimates, simply replace the weight $n^{(k)+}/n^+$ with $N^{(k)}/N$. In other words, weight each stratum by the number of households in that stratum, rather than the number of individuals. While stratification is not necessary to obtain unbiased estimates of household-weighted estimands, stratification will generally improve precision so long as household size is predictive of the outcome.

In practice, it is not always possible or feasible to stratify randomization by household size. Fortunately, researchers can often post-stratify by household size; that is, the researcher can analyze the experiment as if randomization had been stratified by size. In the supplementary materials, we extend the theoretical guarantees from~\citet{miratrix2013adjusting} to two-stage randomization, and include additional discussion of the technical details.

Finally, if there are relatively small samples or the distribution of household sizes varies widely, researchers might want to ``mix and match'' among possible strategies. In the attendance study, there are 3,169 households of size $n_i = 2$ but only two households of size $n_i = 7$. Thus, it is unreasonable to post-stratify precisely on household size. Instead, we post-stratify by dividing household size into $n_i \in \{2, 3, 4-7\}$, using the unbiased IW estimator for households of size four to seven. This is inherently a bias-variance tradeoff and will depend on the particular context. If desired, we could also adjust for $n_i$ via regression, as discussed in Section~\ref{section:covariate-adjustment}~\citep[see also][]{aronow_middleton_cluster}. Of course, researchers should pre-specify such procedures whenever possible.

\section{Variance} \label{section:variance}
We next provide the theoretical variance of the unbiased, weighted 
estimators of Section~\ref{section:unbiased-estimation} as well as a conservative estimator of that variance. We conclude with a brief discussion of inference given a point estimate and its estimated variance.

\subsection{Theoretical and estimated variance}
We give general results for the variance of the unbiased, weighted estimator, of which the household and individual weights are special cases. Given estimand weights $w_i^*$, define the transformed potential outcomes as $Y_{ij}^w(h,z) \equiv N n_i w_i^\ast Y_{ij}(h,z)$. For the household weights, the transformed and original potential outcomes are identical, $Y_{ij}^w(h,z) = Y_{ij}(h,z)$. For the individual weights, the transformed potential outcomes are re-scaled by the relative household size, $Y_{ij}^w(h,z) = (n_i/\bar{n}) \cdot Y_{ij}(h,z)$, where $\bar{n}$ is the average household size. 

We now define several useful terms, 
effectively decomposing the overall variance of the transformed potential outcomes into a within- and between-household variance. 
Let $\sigma^{2,w}_{i,hz} = 1/n_i\sum_j (Y^w_{ij}(h,z) - \overline{Y}^w_i(h,z))^2$ be the within-household potential outcome variances for 
$Y^w_{ij}(h,z)$, and let $\Sigma_{11}^w = \frac{1}{N} \sum_i \sigma^{2,w}_{i,11}$ and 
$\Sigma^w_{10} = \frac{1}{N}\sum_i\frac{1}{(n_i-1)^2}\sigma^{2,w}_{i,10}$ be the (re-scaled) average within-cluster variances for $Y^w_{ij}(1,1)$ 
and $Y^w_{ij}(1,0)$ respectively. Finally, define the between-cluster variance of cluster-level averages:
\begin{eqnarray*}
		V_{hz}^w &=& \frac{1}{N-1} \sum_i^N (\overline{Y}^w_i(h,z) - \overline{Y}^w(h,z))^2, \\
		V_P^w &=& \frac{1}{N-1} \sum_i^N ([\overline{Y}^w_i(1,1) - \overline{Y}^w_i(0,0)] - [\overline{Y}^w(1,1) - \overline{Y}^w(0,0)])^2,\\
		V_S^w &=& \frac{1}{N-1} \sum_i^N ([\overline{Y}^w_i(1,0) - \overline{Y}^w_i(0,0)] - [\overline{Y}^w(1,0) - \overline{Y}^w(0,0)])^2,
\end{eqnarray*}
where $V_{hz}^w$ is the between-cluster variance of the average cluster-level potential outcome, $\overline{Y}^w_i(h,z)$. $V_P^w$ and $V_S^w$ are the (unidentifiable) cluster-level treatment effect variation for the primary and spillover effects, respectively.

\begin{proposition}[Theoretical variance of the two-stage weighted estimators]\label{prop:eq-size-var}
	The two-stage weighted estimators have the following variances under the randomization distribution:
	\begin{equation*}
		\Var(\widehat{\tau}_{W}^P) = \frac{\Sigma_{11}^w + V_{11}^w}{N_1} + \frac{V_{00}^w}{N_0} - \frac{V_P^w}{N}
	\end{equation*}
	and 
	\begin{equation*}
		\Var(\widehat{\tau}_{W}^S) = \frac{\Sigma_{10}^w + V_{10}^w}{N_1} + \frac{V_{00}^w}{N_0} - \frac{V_S^w}{N}.
	\end{equation*}

\end{proposition}
This variance has the same form as the standard Neymanian variance. However, the increased variance due to the two-level randomization is reflected in the first numerator, which has two terms instead of one. Intuitively, this is a decomposition of the marginal variance of potential outcomes into $\Sigma^w_{hz}$, the average of the within-household variances, and $V^w_{hz}$, the variance of the household-level average potential outcomes. 

We can obtain an estimated variance that is a ``conservative'' estimate for the true variance (in the sense of being too wide in expectation) with respect to the randomization distribution. Let $s^{2,w}_{hz}$ be the cluster-level sample variance for the cluster-level average transformed potential outcomes, $\overline{Y}^{\obs,w}_i(h,z)$. That is,
$$s^{2,w}_{hz} = \frac{1}{N_h - 1} \sum_{i}^N \iv(H_i = h)\left(\overline{Y}^{\obs,w}_i(h,z) - \overline{Y}^{\obs,w}(h,z)\right)^2,$$
where $\overline{Y}^{\obs,w}(h,z)$ is the average transformed observed outcome for the set $\mathcal{T}_{hz}$, and where $\overline{Y}^{w,\obs}_i(h,z)$ is the average observed outcome for the set $\mathcal{T}_{hz}$ in household $i$. 
\begin{theorem}[Estimated variance of the two-stage weighted estimators]\label{th:eq-size-var-est-conservative}
	Consider the variance estimators $\Vh(\widehat{\tau}_{W}^P)$ and $\Vh(\widehat{\tau}_{W}^S)$:
	\begin{eqnarray}
	\Vh(\widehat{\tau}_{W}^P) &=& \frac{s^{2,w}_{11}}{N_1} + \frac{s^{2,w}_{00}}{N_0}, \\
	\Vh(\widehat{\tau}_{W}^S) &=& \frac{s^{2,w}_{10}}{N_1} + \frac{s^{2,w}_{00}}{N_0}.
	\end{eqnarray}
	The proposed estimators are conservative estimates of their respective estimands. That is, $\mathbb{E}(\Vh(\widehat{\tau}_{W}^P)) \geq \Var(\widehat{\tau}_{W}^P)$ and $\mathbb{E}(\Vh(\widehat{\tau}_{W}^S)) \geq \Var(\widehat{\tau}_{W}^S)$. $\Vh(\widehat{\tau}_{W}^P)$ and $\Vh(\widehat{\tau}_{W}^S)$ are unbiased if $V_{P}^w = 0$ and $V_{S}^w = 0$, respectively.
\end{theorem}

The results in Proposition~\ref{prop:eq-size-var} and Theorem~\ref{th:eq-size-var-est-conservative} can be applied to HW and IW estimators by simply plugging
the appropriated weights defined in Proposition~\ref{prop:eq-est-unbiasedness}. In particular, plugging in the HW weights recovers the results of \citet{hudgens2012toward}. Consistent with their results, the estimated variance for the weighted estimator is unbiased if the treatment effects are constant. 


\subsection{Inference}
We briefly discuss inference for these quantities given an estimator and its estimated variance, following the setup in~\citet{liu2014large}. For additional discussion, see~\citet{tchetgen2012estimation} and~\citet{Rigdon:2015cv}.

For general outcomes,~\citet{liu2014large} derive both Chebyshev and Wald confidence intervals (CIs) for household weighted estimands under two asymptotic regimes. In the first regime, the number of households (i.e., $N$) remains fixed while the size of each household grows large, i.e., $\text{min}(n_1, \ldots, n_N) \rightarrow \infty$. Inference in this regime either relies on Chebyshev CIs, which are typically too wide to be practically useful, or Wald CIs, which require several additional conditions that are unlikely to hold in our example, including homogeneity of impacts across households.  In the second regime, the size of each hold (i.e., $n$) remains fixed while the number of households grows large, i.e., $N \rightarrow \infty$. At a high level, valid Wald CIs in this regime require a standard Lindeberg condition on the estimators and some restrictions on the within- and between-household variances, as well as requiring that $N_0/N_1$ remains relatively constant as $N\rightarrow \infty$. We rely on this approach here, as it seems like a reasonable asymptotic approximation in our setting. Thus, an asymptotic $1-\gamma$ CI for $\tau^P_{HW}$ is:
$$\widehat{\tau}^P_{HW} \pm z_{1-\gamma/2}\sqrt{ \Vh(\widehat{\tau}^P_{HW}) },$$
where $z_{1-\gamma/2}$ is the $1-\gamma/2$ quantile of the standard Normal distribution. We caution that this asymptotic approximation might have poor performance with a small number of households~\citep[see, for example,][]{clubSandwich_package}. In the supplementary materials we discuss the corresponding assumptions for individual weighted estimands. We argue that we can obtain valid Wald CIs via the $N \rightarrow \infty$ asymptotic regime of~\citet{liu2014large} separately for each household size stratum.

\section{Regression-based estimation}
\label{sect:connection-with-regression}

We now connect these randomization-based results with more familiar regression-based methods. Our key result is that, with the appropriate standard errors, conventional linear regression estimates  are equivalent to the randomization-based estimates. This approach regards regression as a convenient tool~\citep[sometimes known as a \textit{derived linear model}; see][]{hinkelmann2012design}, and does not equate the regression with a specific generative model. In other words, this approach does not impose a model for the potential outcomes. See~\citet{baird2014designing} for additional discussion of regression for two-stage randomized designs.

We consider two basic regression approaches, an individual-level regression and a household-level regression. For simplicity, we start with the equal-sized household case and then show that these results generalize to any two-stage weights. We then demonstrate the dangers of using standard errors that ignore the two-stage structure. This section builds on existing results for robust and cluster-robust standard errors, especially~\citet{McCaffrey2001generalizations} and~\citet{Bell2002bias}. See also~\citet{cameron2015practitioner, imbens2012robust, clubSandwich_package}.

\subsection{Individual-level regression}\label{section:individual-level-regression}
First, we construct the individual-level linear model,
\begin{equation}\label{eq:equal-linear-model-bad}
	Y_{ij}^{\obs} = \alpha + \beta^P H_i Z_{ij} + \beta^S H_i (1 - Z_{ij}) + \varepsilon_{ij},
\end{equation}

where the uncertainty in $\varepsilon_{ij}$ is entirely due to randomization. It is straightforward to show that standard OLS estimates for $\beta^P$ and $\beta^S$ are identical to the randomization-based estimators in Section~\ref{section:unbiased-estimation}. The theorem below states that the the randomization-based standard errors are equivalent to a particular cluster-robust generalization of heteroskedasticity-consistent standard errors, known as \textit{HC2}~\citep{mackinnon1985some}.

\begin{theorem}\label{th:reg-cluster-robust} For equal-sized households, let $Y$ be the vector containing all the observed outcomes $Y^\obs_{ij}$; let $X$ denote the appropriate design matrix (formally defined in the supplementary materials) with columns corresponding to the intercept, $HZ$, and $H(1-Z)$; and let $\beta = (\alpha, \beta^P, \beta^S)$. The linear model in Equation~\ref{eq:equal-linear-model-bad} can therefore be re-written as $Y = X\beta + \varepsilon$, with corresponding least squares estimate, $\widehat{\beta}^{\ols}$. These estimates are unbiased for their corresponding estimands.
	Further, define the cluster-robust generalization of HC2 standard errors as:
	\begin{equation*}
		\widehat{Var}^{clust}_{hc2}(\widehat{\beta}^{\ols}) = (X^tX)^{-1} \sum_{s=1}^S X_s^t(I_{N_s} - P_{ss})^{-1/2} \widehat{\varepsilon}_s \widehat{\varepsilon}_s
		(I_{N_s} - P_{ss})^{-1/2} X_s (X^tX)^{-1}
	\end{equation*}
	where $X_s$ and $\widehat{\varepsilon}_{s}$ are the subsets of $X$ and $\widehat{\varepsilon}$ corresponding to household $s$, and $P_{ss}$ is
	defined as $P_{ss} = X_s(X^tX)^{-1} X_s^t.$
	Then 
$$		\widehat{Var}^{clust}_{hc2}(\widehat{\beta}^{P,\ols}) = \widehat{Var}(\widehat{\tau}^P) \,\,\,\,\,\, \mbox{ and } \,\,\,\,\,\,
		\widehat{Var}^{clust}_{hc2}(\widehat{\beta}^{S,\ols}) = \widehat{Var}(\widehat{\tau}^S)$$
where $\widehat{Var}^{clust}_{hc2}(\widehat{\beta}^{P,\ols}) = (\widehat{Var}^{clust}_{hc2}(\widehat{\beta}^{\ols}))_{22}$ and $\widehat{Var}^{clust}_{hc2}(\widehat{\beta}^{S,\ols}) = \widehat{Var}^{clust}_{hc2}(\widehat{\beta}^{ols}))_{33}$.

\end{theorem}

In short, Theorem~\ref{th:reg-cluster-robust} confirms that we can obtain the same randomization-based point- and variance-estimators via the individual-level linear model in Equation~\ref{eq:equal-linear-model-bad} with HC2 cluster-robust standard errors. This is similar to results obtained with heteroskedastic-robust standard errors in simpler designs~\citep[e.g.,][]{samii2012equivalencies,imbens2015causal}. In Section~\ref{section:no_clustering}, we demonstrate the effect of failing to account for clustering on standard errors. Researchers can estimate these standard errors directly in \texttt{R} via, for example, the \texttt{clubSandwich} package. See~\citet{clubSandwich_package} for additional discussion on the performance of clustered standard errors with a relatively small number of clusters.

\subsection{Household-level regression}
\label{section:cluster-level-regression}
We now consider regression at the household level. This is a common strategy in cluster-randomized trials and yields identical inference to individual-level regression with clustered standard errors~\citep[see, e.g.,][]{cameron2015practitioner, Athey:2016wn}.

We separately aggregate treated and control units within each treated household, thus considering three types of household-level aggregates for household $i$. Each treated household has two household-level averages, $\overline{Y}_i^\obs(1,1)$ and $\overline{Y}_i^\obs(1,0)$; each control household has one household-level average, $\overline{Y}_i^\obs(0,0)$. We can therefore assemble a vector of household-average outcomes, $\overline{Y}_k^\obs$ of length $2N_1 + N_0$. We introduce the indicators $H^{(11)}_k$ and $H^{(10)}_k$; $H_k^{(11)} = 1$ if the aggregate is over the 
treated units in treated households and $H_k^{(11)} = 0$ otherwise; $H_k^{(10)} = 1$
if the aggregate is over the control units in treated households and $H_k^{(10)} = 0$ otherwise.
We then consider the following linear model:
\begin{equation}\label{eq:equal-linear-model-good}
	\overline{Y}_k^{\obs} = \alpha + \beta^P H_k^{(11)} + \beta^S H_k^{(10)} + \varepsilon'_k.
\end{equation}

We now show that we can obtain the randomization-based point and variance estimates via the linear model estimates with standard (i.e., non-cluster) heteroskedastic-robust standard errors. 
\begin{theorem}\label{th:reg} For equal-sized households, the OLS estimates for $\beta^P$ and $\beta^S$ in Equation~\ref{eq:equal-linear-model-good} are unbiased estimators of the corresponding estimands. Define the heteroskedastic-robust estimator of the variances: 
	\begin{equation*}
		\Vhc(\widehat{\tau}^P) \equiv \Vhc(\widehat{\beta}^P) = \frac{\sum_{k: H^{(10)}_k=0} \widehat{\varepsilon}_{k}^2 (H^{(11)}_k -\overline{H})^2 }{(\sum_{k: H_k^{(10)}=0}(H^{(11)}_k - \overline{H}^{(11)})^2)^2}
	\end{equation*}
	and
	\begin{equation*}
		\Vhc(\widehat{\tau}^S) \equiv \Vhc(\widehat{\beta}^S) = \frac{\sum_{k: H^{(11)}_k=0} \widehat{\varepsilon}_{k}^2 (H^{(10)}_k -\overline{H})^2 }{(\sum_{k: H_k^{(11)}=0}(H^{(10)}_k - \overline{H}^{(10)})^2)^2}
	\end{equation*}
	We have:
	\begin{equation*}
		\Vhc(\widehat{\tau}^P)  = \Vh(\widehat{\tau}^P) \,\,\,\,\,\, \mbox{ and } \,\,\,\,\,\, \Vhc(\widehat{\tau}^S)  = \Vh(\widehat{\tau}^S),
	\end{equation*}
	where the $\widehat{\varepsilon}_{k}$'s are the $HC2$ residuals (see supplementary materials for exact definition).
	\end{theorem}
	
In short, Theorem~\ref{th:reg} states that we can aggregate to the household level and proceed as if this were a standard completely randomized trial~\citep{imbens2015causal}.
Intuitively, the aggregation at the household level is another way of accounting for the household structure in the two-stage randomization scheme. Since the definition for the heteroskedastic-robust estimator of the variance is standard, it is straightforward to fit Equation~\ref{eq:equal-linear-model-good} and the corresponding variance in \texttt{R} via, for example, the \texttt{vcovHC} function with the \texttt{HC2} option in the \texttt{sandwich} package~\citep{zeileis2004sandwich}.

\subsection{Results for weighted estimands}
Finally, we can modify the household-level regression to estimate any two-stage weighted estimand. Thus, we can use this approach to recover the results in Sections~\ref{section:estimation} and \ref{section:variance}. The key idea is to run an unweighted regression on transformed outcomes. 

\begin{theorem}\label{th:reg-weight}
	Let $w_i^*$ be two-stage estimand weights, with 
	transformed potential outcomes, $Y^{obs,w}_{ij}(h,z) = N n_i w_i^* Y^{obs}_{ij}(h,z)$. Then the results of 
	Theorem~\ref{section:cluster-level-regression} hold when applied to the transformed potential outcomes. That is: $\widehat{\beta}^P = \widehat{\tau}_W^P$,  $\widehat{\beta}^S = \widehat{\tau}_W^S$, $\widehat{Var}_{hc2}(\widehat{\beta}^P) = \widehat{Var}(\widehat{\tau}_W^P)$, and $\widehat{Var}_{hc2}(\widehat{\beta}^S) = \widehat{Var}(\widehat{\tau}_W^S)$.
	%
\end{theorem}

\noindent This approach is subtly different from using Weighted Least Squares, which would also reweight the design matrix.

\subsection{Failing to account for clustering}
\label{section:no_clustering}

It is instructive to consider the consequences of ``naively'' analyzing a two-stage experiment as if it were a completely randomized experiment, ignoring the household structure. With equal-sized households, the point estimates will be the same as for the appropriate analysis, but the standard errors will differ. In particular, let $\widehat{\Var}^{\text{het}}_{\text{hc2}}(\widehat{\tau}^P)$ be the (non-cluster) HC2 robust standard error for the primary effect; that is, these are the variances from Equation (\ref{eq:equal-linear-model-good}) for the household aggregates incorrectly applied to the individual level. We show in the supplementary materials that, 
\begin{equation*}
\mathbb{E}\bigg[\widehat{\Var}^{\text{het}}_{\text{hc2}}(\widehat{\tau}^P)\bigg] - \Var(\widehat{\tau}^P)=(\Sigma_{00} + V_{00})\left\{\frac{1}{nN_0-1}\bigg(1 - n\rho_{00}\bigg) - \frac{1}{N}\frac{V_p}{\Sigma_{00} + V_{00}}\right\},
\end{equation*}
where $V_{00}$ and $\Sigma_{00}$ are the between- and within-household variances, respectively, of the control potential outcomes, $Y_{ij}(0,0)$, and $\rho_{00} \equiv V_{00}/(\Sigma_{00} + V_{00})$ is the intraclass correlation (ICC).
This quantity is negative---that is, the variance is anti-conservative in expectation---if:
$$\rho_{00} > \frac{1}{n} - \left(\frac{nN_0 - 1}{nN}\right) \left(\frac{ V_P }{\Sigma_{00} + V_{00}}\right).$$
Since the last term is non-negative, we can build intuition by setting that term to zero. For example, consider the special case of $V_P = 0$, which would occur if there is a constant additive effect. Under these conditions, the estimated variance is anti-conservative if $\rho_{00} > 1/n$. In the social sciences, typical values of ICC range from 0.1 to 0.3~\citep[e.g.,][]{gelman2006data}. Thus, even with households of size 4 or 5, the estimated variance could be anti-conservative. We see this behavior in the simulations in Section~\ref{section:simulations}.

\section{Covariate adjustment}
\label{section:covariate-adjustment}

Finally, we explore how to incorporate individual- and cluster-level covariates in a two-stage experiment. There is an extensive literature on the use of covariates in randomized trials~\citep[see, for example,][]{imbens2015causal}. In this section, we briefly address stratification, post-stratification, and model-assisted estimation. 

\paragraph{Stratification and post-stratification.} As with household size in Section~\ref{section:stratified-randomization}, the simplest way to account for covariates is to incorporate them into the randomization by stratifying on them. In general, the researcher can partition households into discrete strata, regard each stratum as a separate ``mini experiment,'' and estimate a pooled effect by averaging across strata. This will improve the precision of the treatment effect estimate so long as the stratifying covariate is predictive of the outcome. Researchers cannot, however, stratify by a covariate that varies within household, as this could destroy the nested structure in the data by assigning different individuals in the same household to different ``household-level'' treatments. For example, we cannot stratify the two-stage randomization by gender, as some houses will have both boys and girls; instead, we could stratify by whether all students in the household are boys, which is an aggregate version of the individual-level covariate. Finally, as with household size, researchers can also post-stratify on household-level covariates. Of course, it is possible to combine stratification and post-stratification; for example, first stratify by household size and then, for each household size, post-stratify by whether the household speaks English as the primary language.

\paragraph{Model-assisted estimation.} We also consider model-assisted estimation to incorporate covariates~\citep{cochran::1977, rosenbaum2002covariance, hansen2009attributing, aronow2013class}. 
Following the setup in~\citet{hansen2009attributing}, consider $K$ covariates $x^{(1)}, \ldots, x^{(K)}$ (which typically include a constant) with corresponding coefficient vector, $\gamma = (\gamma_1, \ldots, \gamma_K)$, such that  $r(\gamma) = \{r_{ij}(\{x^{(k)}\}_k, \gamma)\}$ for $i=1,\ldots,N$ and $j=1,\ldots,n_i$ is a function mapping covariates to predictions. To simplify notation, we will let $x = \{x^{(k)}\}_k$. In practice, the coefficients in $\gamma$ are typically coefficients from a linear regression. In this case, let 
\begin{equation*}
	r_{ij}(x, \gamma) = \sum_k^K x_{ij}^{(k)} \gamma_k,
\end{equation*}
where $x_{ij}$ is a vector of covariates associated with unit $j$ in cluster $i$. We regard $r_{ij}(x, \gamma)$ as fixed and known for all units, rather than estimated from the data. We then define the (residualized) potential outcome as:
\begin{equation}
	e^\gamma_{ij}(h,z) = Y_{ij}(h,z) - r_{ij}(x, \gamma).
\end{equation}
As with the corresponding potential outcomes, $Y_{ij}(h,z)$, the residualized potential outcomes, $e^\gamma_{ij}(h,z)$, are assumed to be fixed and are only observed if $H_i = h$ and $Z_{ij} = z$. 
We can then substitute $e^\gamma_{ij}(h,z)$ for $Y_{ij}(h,z)$ in defining the primary effect:
 \begin{align*}
	\tau^P &= \overline{Y}(1,1) - \overline{Y}(0,0) \\
	&= \left[\overline{e}^\gamma(1,1) + \overline{r}(x, \gamma)\right] - \left[\overline{e}^\gamma(0,0) + \overline{r}(x, \gamma)\right] = \overline{e}^\gamma(1,1) - \overline{e}^\gamma(0,0),
\end{align*}
re-writing $\tau^S$ in an analogous way.  

Given $r_{ij}(\gamma)$, model-assisted estimation of $\tau^P$ is immediate via substituting the observed values of the residualized outcomes, $e^{\gamma, \obs}_{ij}$, in place of the unadjusted outcomes, $Y_{ij}^\obs$. The resulting difference-in-means estimator is unbiased regardless of the exact values of $r_{ij}(\gamma)$; that is, there is no need to appeal to a ``correctly specified'' linear model to obtain the particular coefficient vector, $\gamma$. So long as the covariates are predictive of the outcome, the variance of $e^{\gamma, \obs}_{ij}$ will generally be smaller than the variance of $Y^\obs_{ij}$, and the resulting model-assisted estimator will also have smaller estimated variance (so long as $\gamma$ does not include extreme values). 

Finally, the above derivations assume that $\gamma$ is fixed and known; in practice, we must find some way to determine $\gamma$. The most straightforward approach is to generate a random hold-out sample, and estimate $\gamma$ via a regression of $Y$ on $X$ for this group. While not always possible, the attendance study effectively has a hold-out sample that we use for this purpose; see Section~\ref{section:data}. See~\citet{hansen2009attributing} and~\citet{aronow2013class} for additional discussion. 



\section{Simulations}
\label{section:simulations}

\subsection{Failing to account for the cluster structure}

We now turn to simulations assessing the importance of accounting for the cluster structure. 
Reflecting the nested structure in the actual experiment, we generate potential outcomes in two stages. First, we simulate household-level average potential outcomes via:
\begin{align*}
	\overline{Y}_{i}(0,0) &\sim N\left(\mu_{00}, \sigma^2_{\mu}\right) \\
	\tau_i^P &\sim N\left(\overline{\tau}^P, \sigma^2_{\tau^P}\right)\\
	\tau_i^S &\sim N\left(\overline{\tau}^S, \sigma^2_{\tau^S}\right),
\end{align*}
where $\overline{Y}_{i}(1,1) = \overline{Y}_{i}(0,0) + \tau_i^P$ and $\overline{Y}_{i}(1,0) = \overline{Y}_{i}(0,0) + \tau_i^S$. Then, conditional on these values, we generate individual-level potential outcomes via
$Y_{ij}(h,z) \sim N\left(\overline{Y}_i(h,z), \sigma^2_y\right),$ for each $h$ and $z$. Across all simulations, we fix the mean potential outcomes, with $\mu_{00} = 2$, $\bar{\tau}^S = 0.7$, and $\bar{\tau}^P = 1.5$, and fix household size at $n_i = 4$ for all households. For convenience, we also restrict the household-level variance terms to be equal to each other, such that $\sigma_\mu = \sigma_{\tau^P} = \sigma_{\tau^S} = \sigma_c$, where $\sigma_c$ is the common standard deviation. Thus, we vary three main parameters, $N \in \{50, 100, 500, 1000\}$ (we always set $N_1=N/2$) and $\sigma_{c}, \sigma_y \in \{0.1, 0.2, 0.3, 0.4, 0.5\}$. For each combination of parameters, we consider three methods: cluster-robust standard errors; non-cluster, robust standard errors; and nominal standard errors. For each, we compute the coverage of the associated 95\% confidence interval, averaging over 2,000 random draws of the assignment vector.

\begin{table}[btp]
	\centering
	\caption{Average coverage of 95\% confidence intervals over 2,000 replications.}
	\label{tbl:sim_overview}
	\begin{tabular}{r cc}
	& $\widehat{\tau}^P$ & $\widehat{\tau}^S$ \\
	\hline
Cluster robust SEs & 0.97 & 0.98 \\
Non-cluster robust SEs & 0.93 & 0.87 \\
Nominal SEs & 0.95 & 0.86 \\
\hline		
	\end{tabular}

\end{table}

\begin{figure}[btp]
	\begin{subfigure}[b]{0.5\textwidth}
	\centering
	\includegraphics[width=\textwidth]{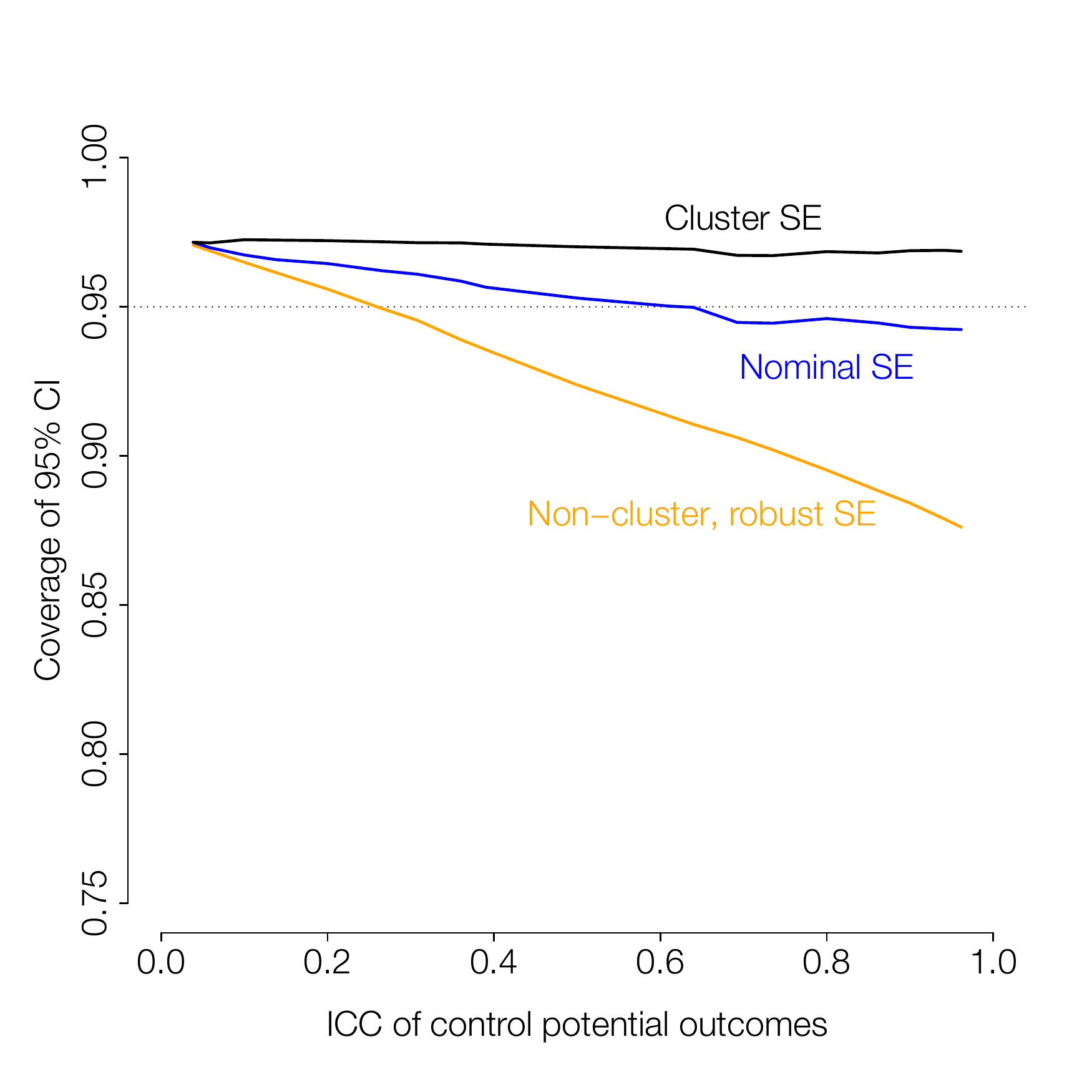}
	\caption{Primary effect}
	\label{fig:coverage_primary}
	\end{subfigure}%
	\begin{subfigure}[b]{0.5\textwidth}
	\centering
	\includegraphics[width=\textwidth]{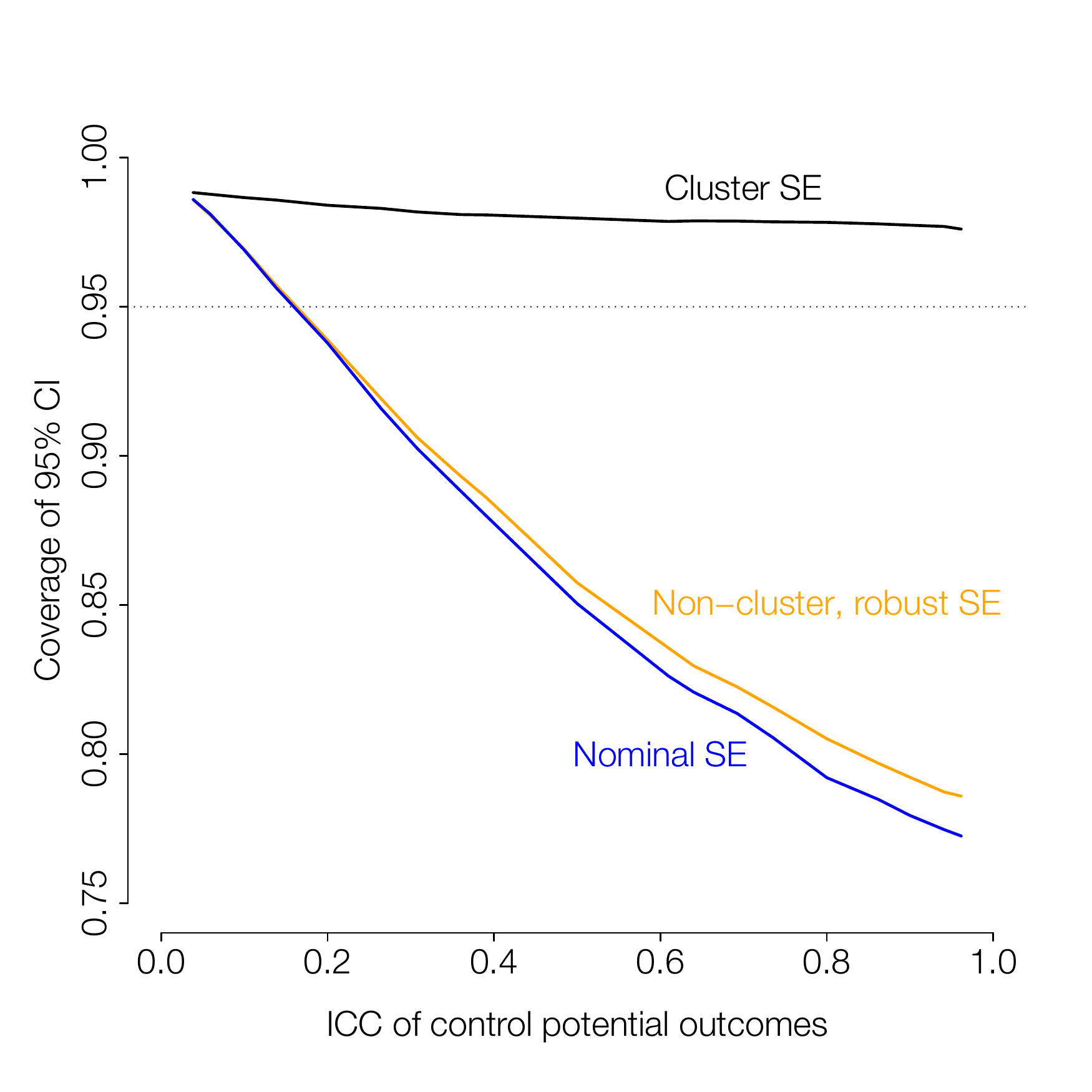}
	\caption{Spillover effect}
	\label{fig:coverage_spillover}
	\end{subfigure}
	\caption{Coverage for 95\% confidence intervals for clustered SEs, non-cluster robust SEs, and nominal SEs, with respect to the intraclass correlation of control potential outcomes, $\sigma^2_c/(\sigma^2_c + \sigma^2_y)$.}
	\label{fig:coverage_icc}
\end{figure}

Table~\ref{tbl:sim_overview} shows the overall coverage for 95\% confidence intervals, averaged across all values of the simulation parameters. As expected, coverage with the cluster-robust standard errors is slightly larger than 95\% coverage. By contrast, the non-cluster and nominal standard errors have below 95\% coverage. First, this coverage pattern is quite stable across different sample sizes. At the same time, coverage strongly depends on the within- and between-household variances. Figure~\ref{fig:coverage_icc} shows the relationship between coverage and the intraclass correlation among control potential outcomes, defined as $\sigma^2_c/(\sigma^2_c + \sigma^2_y)$ in this simulation. Consistent with the results in Section~\ref{section:no_clustering}, the coverage for non-clustered standard errors grows increasingly poor as the ICC increases.

\subsection{Comparing the three estimators for the IW estimand}

We now focus on the IW estimand and compare the unbiased estimator, the difference-in-means estimator,
and the post-stratified estimator. We consider two scenarios: (a) when treatment effects are uncorrelated with household size; and (b) when treatment effects are correlated
with household size. The data generating process is the same as above, with a balanced household-level randomization with $N = 200$, $N_1 = 100$, and fixed 
$\sigma_c = \sigma_y = 0.3$. We generate households of size 2, 3 and 4 with equal probability, and introduce
the parameters $\overline{\tau}^P_k, \overline{\tau}^S_k, \mu_{00}^{(k)}$ for $k = 2,3,4$. For scenario (a), we set
$\overline{\tau}^P_k = 1.5, \overline{\tau}^S_k=0.7, \mu_{00}^{(k)}=2$ for all $k=2,3,4$. For scenario (b), we allow
the effects to vary by household size, as follows: $\overline{\tau}^P_2 = 1.5, \overline{\tau}^P_3 = 0.75, 
\overline{\tau}^P_4 = 0.37$, $\overline{\tau}^S_2 = 0.7, \overline{\tau}^S_3 = 0.35, \overline{\tau}^S_4 = 0.17$, 
and $\mu_{00}^{(2)} = 2, \mu_{00}^{(3)}=1, \mu_{00}^{(4)} = 0.5$. 

The results are presented in Table~\ref{table:iw-results}. We see that when the treatment effect is uncorrelated with household size, the bias of all three estimators is negligible, but the Monte Carlo standard error of the unbiased estimator is an order of magnitude larger than that of the other two estimators. When treatment effect is correlated with household size, the biases of the unbiased and the post-stratified estimators are still very small, but the bias of the simple difference estimator is substantial---roughly the same size as the standard error. Again, the standard errors are smallest for the post-stratified estimator; overall, the post-stratified estimator clearly dominates in terms of RMSE.

\begin{table}
	\centering
	\caption{Bias and SE for different estimators for the IW estimand over 2,000 replications. `\textemdash' denotes $\leq 10^{-3}$.}
	\label{table:iw-results}
	\begin{tabular}{ll cc c cc}
		& & \multicolumn{2}{c}{hh size uncorrelated with effect} & & \multicolumn{2}{c}{hh size correlated with effect} \\
		\cline{3-4} \cline{6-7}
		\multicolumn{2}{l}{Estimator} & Avg. $|$bias$|$ & Monte Carlo SE & & Avg. $|$bias$|$ & Monte Carlo SE \\
		\hline
		\multicolumn{2}{l}{\textit{Primary effect}} & &&&& \\
		& Unbiased & \textemdash & 0.11 & & \textemdash & 0.07 \\
		& Simple difference &  \textemdash & 0.04 & & 0.18 & 0.12 \\
		& Post-stratified &  \textemdash & 0.04 & &\textemdash & 0.04 \\[1em]
		\multicolumn{2}{l}{\textit{Spillover effect}} & &&&& \\
		& Unbiased & \textemdash & 0.12 & & \textemdash & 0.07\\
		& Simple difference & \textemdash & 0.04 & &  0.12 & 0.13 \\
		& Post-stratified & \textemdash & 0.04 & & \textemdash & 0.04  \\
		\hline
	\end{tabular}
\end{table}


\section{Student absenteeism in the School District of Philadelphia}
\label{section:data}

\subsection{Setup and covariate balance}
We now apply these methods to our motivating example, the~\citet{Rogers_Feller_SDP} attendance intervention in the School District of Philadelphia. The original study included three active treatment arms and one control arm, with $N = 28,080$ total households. The first treatment arm merely reminded parents of the importance of attendance and did not provide any student-specific information. The second and third arms provided parents with different types of student-specific information.~\citet{Rogers_Feller_SDP} found weak impacts of assignment to the first arm and large impacts for the second and third arms (all relative to control). They also found minimal differences between the second and third arms.

We now make three modifications that preserve the substantive questions from the original study but allow us to focus on the two-stage randomized design. First, we exclude households from the weak first treatment arm, instead using this group as the holdout sample for estimating the covariate adjustment model (see Section~\ref{section:covariate-adjustment}). Second, we combine the second and third treatment arms, since these are substantively very similar and have virtually identical impacts. Together, these modifications essentially create a two-arm trial: control vs. combined second and third arms from the original study. Finally, we restrict the universe to the subset of households with two or more eligible students, which yields $N = 3,876$ total households, of which $N_1 = 2,568$ (66 percent) were assigned to treatment and $N_0 = 1,308$ (34 percent) were assigned to control, and $n^+ = 8,654$ total students.

We next assess covariate balance for this sample.  While household-level randomization was not stratified by household size (see Section~\ref{section:stratified-randomization}), the balance by household size is excellent, as shown in Table~\ref{tbl:num_students_hh}. Table~\ref{tbl:cov_balance} shows the covariate balance for each stage of randomization. The left bank shows balance for the first stage, household-level randomization, with covariates aggregated to the household level. The right bank shows balance for the second stage randomization, with individual-level covariates among households assigned to treatment ($H_i = 1$). Statistically, Table~\ref{tbl:cov_balance} shows that covariate balance is excellent for both stages of randomization, with all normalized differences~\citep{imbens2015causal} below 0.05 in absolute value. Substantively, Table~\ref{tbl:cov_balance} emphasizes that the students come from largely disadvantaged households. Over three-quarters of these students qualify for Free or Reduced Price Lunch, which is only available to families at or near the Federal Poverty Line. Over 15 percent of households speak a language other than English at home, with 7 percent of students designated as Limited English Proficiency (LEP). Moreover, this is a very high-absence group, with an average of around 13 days absent in the previous school year (out of roughly 180 possible days); we also include number of absences prior to randomization in early October. We observe the grade for each student, which we treat as a discrete covariate and which ranges from first grade to high school senior. While we do not show balance by grade to conserve space, there is excellent balance across this covariate as well.

\begin{table}[btp]
	\centering
	\caption{Covariate means and normalized differences ($\Delta$) by stage of randomization. For the first stage, balance is assessed for covariates aggregated to the household level. For the second stage, balance is assessed for individual-level covariates, restricted to households assigned to treatment ($H_i = 1$).}
	\label{tbl:cov_balance}
	\begin{tabular}{l ccc c ccc}
\hline
& \multicolumn{3}{c}{\textit{First Stage} (all households)} & & \multicolumn{3}{c}{\textit{Second Stage} ($H_i = 1$ only)} \\
& \multicolumn{3}{c}{Household-level means} & & \multicolumn{3}{c}{Individual-level means} \\
\cline{2-4} \cline{6-8}
& For $H_i = 1$ & For $H_i = 0$ & $\Delta$ & & For $Z_{ij} = 1$ & For $Z_{ij} = 0$ & $\Delta$ \\
\cline{2-4} \cline{6-8}
Female & 0.53 & 0.54 & -0.03 & & 0.53 & 0.53 & -0.01\\
Black/African-American & 0.51 & 0.50 & 0.03 & & 0.52 & 0.52 & -0.01\\
English spoken at home & 0.84 & 0.83 & 0.04 & & 0.84	 & 0.83 & 0.02 \\
Limited English Proficiency & 0.07 & 0.07 & -0.01 & & 0.07 & 0.08 & -0.04\\
Free or Reduced Price Lunch & 0.78 & 0.79 & -0.03 & & 0.78 & 0.79 & -0.02\\
Prior year absences (days) & 16.70 & 16.55 & 0.02 & & 16.65 & 16.93 & -0.03\\
Start-of-year absences (days) & 1.20 & 1.14 & 0.05 & & 1.22 & 1.21 & 0.01\\
Students per household ($n_i$) & 2.2 & 2.2 & -0.01 & & \multicolumn{3}{c}{\textemdash}\\
\hline
\end{tabular}
\end{table}

Finally, we are broadly interested in days absent as the outcome of interest. However, the distribution of absences has a long right tail; for example, several students in the sample are absent for over half the school year. As this greatly increases the variance, we consider two transformed outcomes of interest. First, we consider an indicator for whether a student is chronically absent, defined as missing 18 or more days during the school year, i.e., $\iv(\text{days} \geq 18)$; among students in the control group, 36 percent are chronically absent. Second, we consider log-absences, defined as $\text{log}(\text{days} + 1)$, to allow for a continuous outcome without the very heavy right tail; baseline absences among students in the control group are around 13 days or $\text{log}(13 + 1) \approx 2.6$. For interpretability, we also report key point estimates in terms of raw days. 


\subsection{Results}
\begin{figure}[btp]
	\begin{subfigure}[b]{0.5\textwidth}
		\centering
		\includegraphics[width = \textwidth]{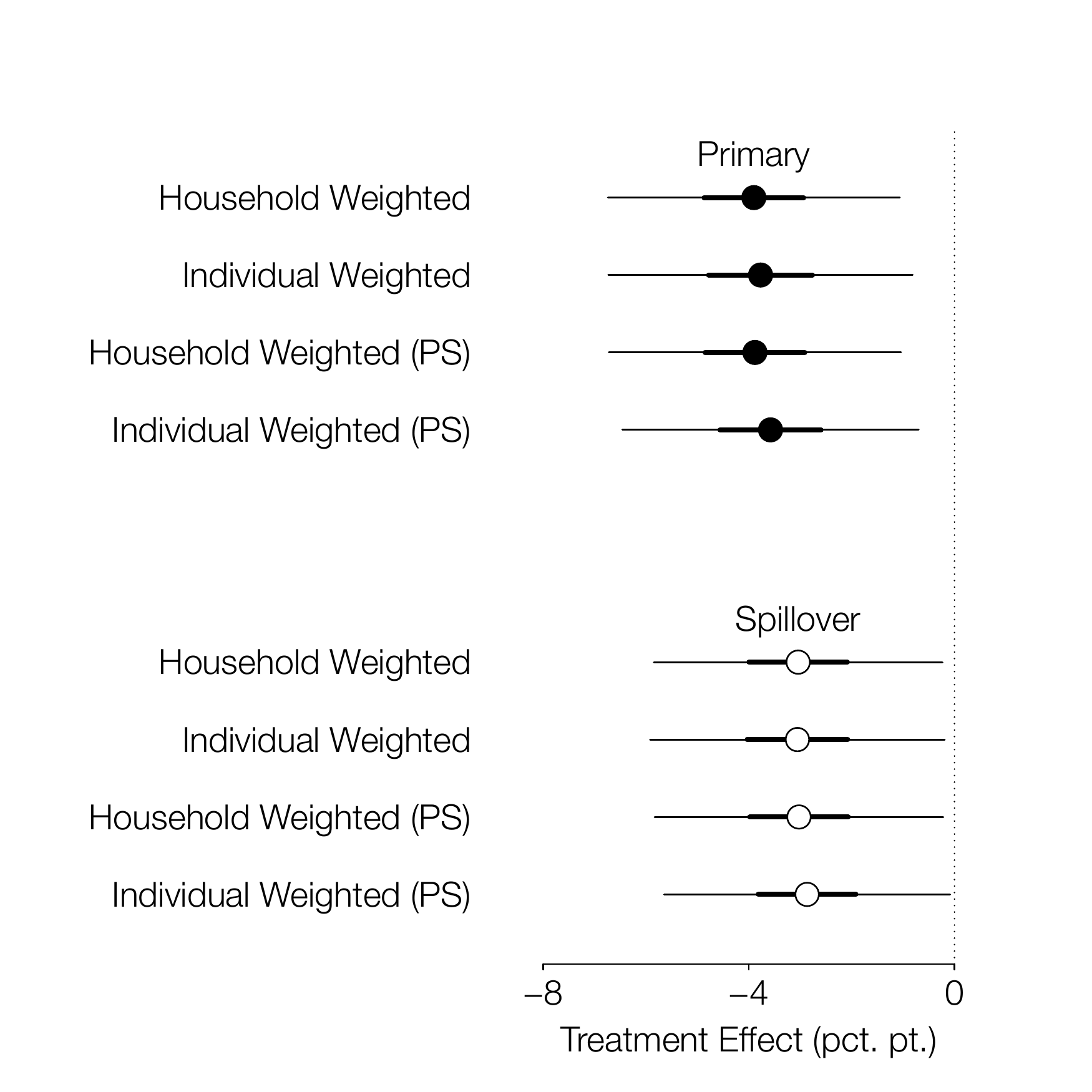}	
		\caption{Binary outcome: chronically absent}
	\end{subfigure}%
	~\begin{subfigure}[b]{0.5\textwidth}
		\centering
		\includegraphics[width = \textwidth]{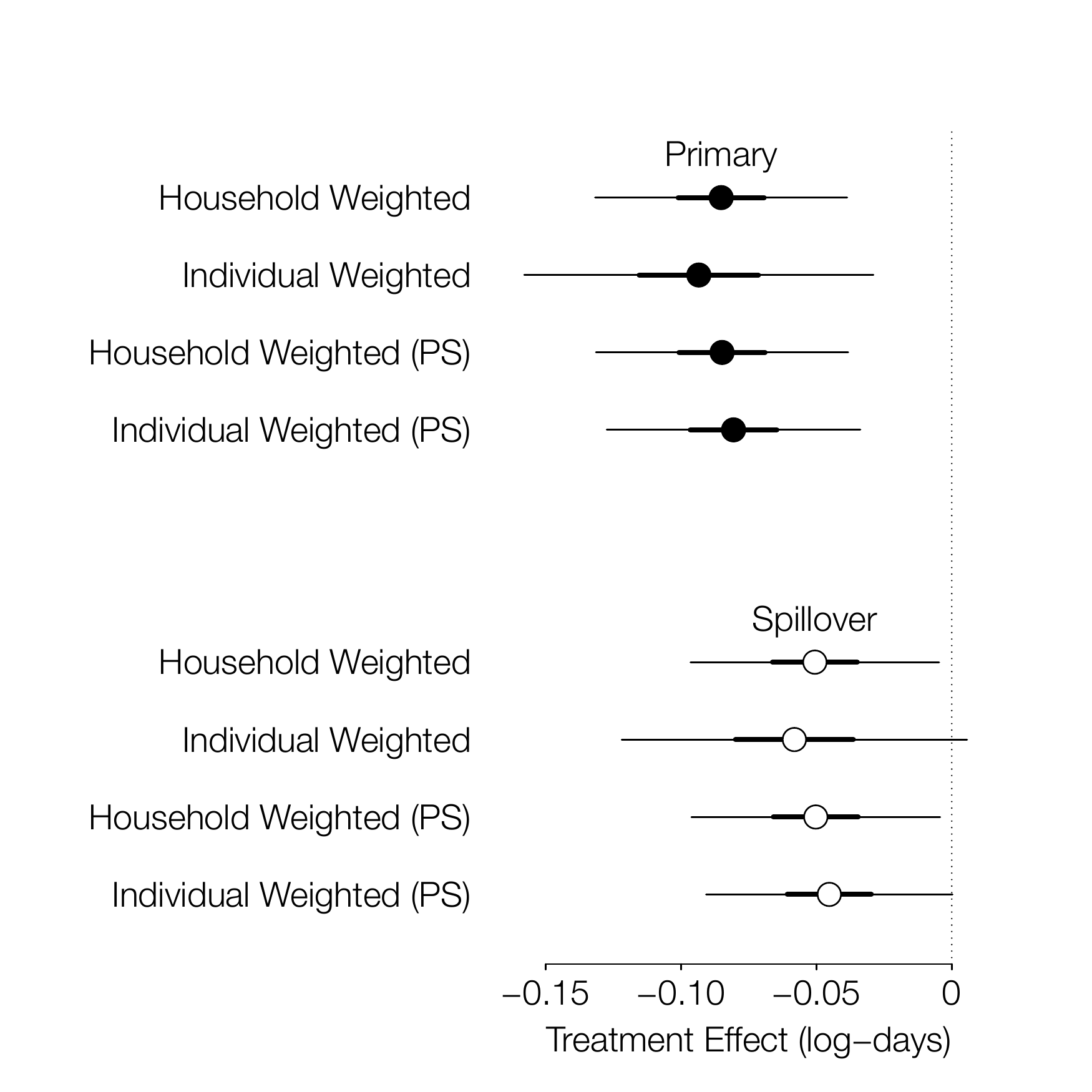}	
		\caption{Continuous outcome: log(Days + 1)}
	\end{subfigure}
	\caption{Treatment effect estimates and 95\% confidence intervals for primary (filled-in circles) and spillover (open circles) effects, for household- and individual-weighted estimands with and without post-stratification (PS) by household size.}
	\label{fig:SDP_main_effects}	
\end{figure}

Figure~\ref{fig:SDP_main_effects} shows the estimated impacts and corresponding 95\% confidence intervals for the primary and spillover effects for both household- and individual-weighted estimands. In terms of chronic absenteeism (i.e., the binary outcome), the estimates for the HW and IW estimands are nearly identical: the unbiased estimates for the primary effects are around -4 percentage points (SE of 1.5 percentage points) for both $\tau_{HW}^P$ and $\tau_{IW}^P$;  the unbiased estimates for the spillover effects are around -3 percentage points (SE of 1.5 percentage points) for both $\tau_{HW}^S$ and $\tau_{IW}^S$. These results are virtually unchanged when post-stratifying by household size, using the (conditionally) unbiased estimator within each post-stratification cell, defined by $n_i = 2$, $n_i = 3$, and $n_i \in \{4, \ldots, 7\}$. As discussed in Section~\ref{section:overall-effect} and in the supplementary materials, the overall effect is essentially a weighted average of the primary and spillover effects. Unsurprisingly, the unadjusted estimate of the overall effect is around 3.5 percentage points (SE of 1.4 percentage points) for both the HW and IW weighted estimands, with nearly identical results for the post-stratified estimate.

The results are somewhat more variable for the impact on log-absences (i.e., the continuous outcome). The point estimates are quite close for the household-weighted and individual-weighted estimands: $\widehat{\tau}^P_{HW} = -0.085$ log-days and $\widehat{\tau}^P_{IW} = -0.093$ log-days for the primary effect, and  $\widehat{\tau}^S_{HW} = -0.051$ log-days and $\widehat{\tau}^S_{IW} = -0.058$ log-days for the spillover effect. In terms of raw days, these are roughly -1.2 days for the primary effect and -0.7 days for the spillover effect. The point estimates are similarly close for the post-stratified estimator. The standard errors, however, are considerably larger for the unadjusted IW estimates: roughly 0.033 log-days for the IW estimands compared to 0.023 log-days for the HW estimands. Thus, the corresponding confidence intervals are roughly 50 percent larger for the IW estimands than for the HW esitmands. Post-stratification greatly reduces the standard errors for the IW estimand: for both IW and HW estimates, the standard errors are roughly 0.023 log-days, comparable to the standard errors for the HW estimate without post-stratification. As with the impact on chronic absenteeism, the estimates for the overall effect fall between the primary and spillover effect estimates, with estimates around -0.07 log-days (SE of 0.023 log-days) or around 1 day.

Despite minor differences between the sets of estimates, the pattern of effects is fairly clear: in general, we find that the spillover effect is between 60 and 80 percent as large as the primary effect, depending on the outcome. We also find few differences between the HW and IW estimates.

\begin{figure}[btp]
	\begin{subfigure}[b]{0.5\textwidth}
		\centering
		\includegraphics[width = \textwidth]{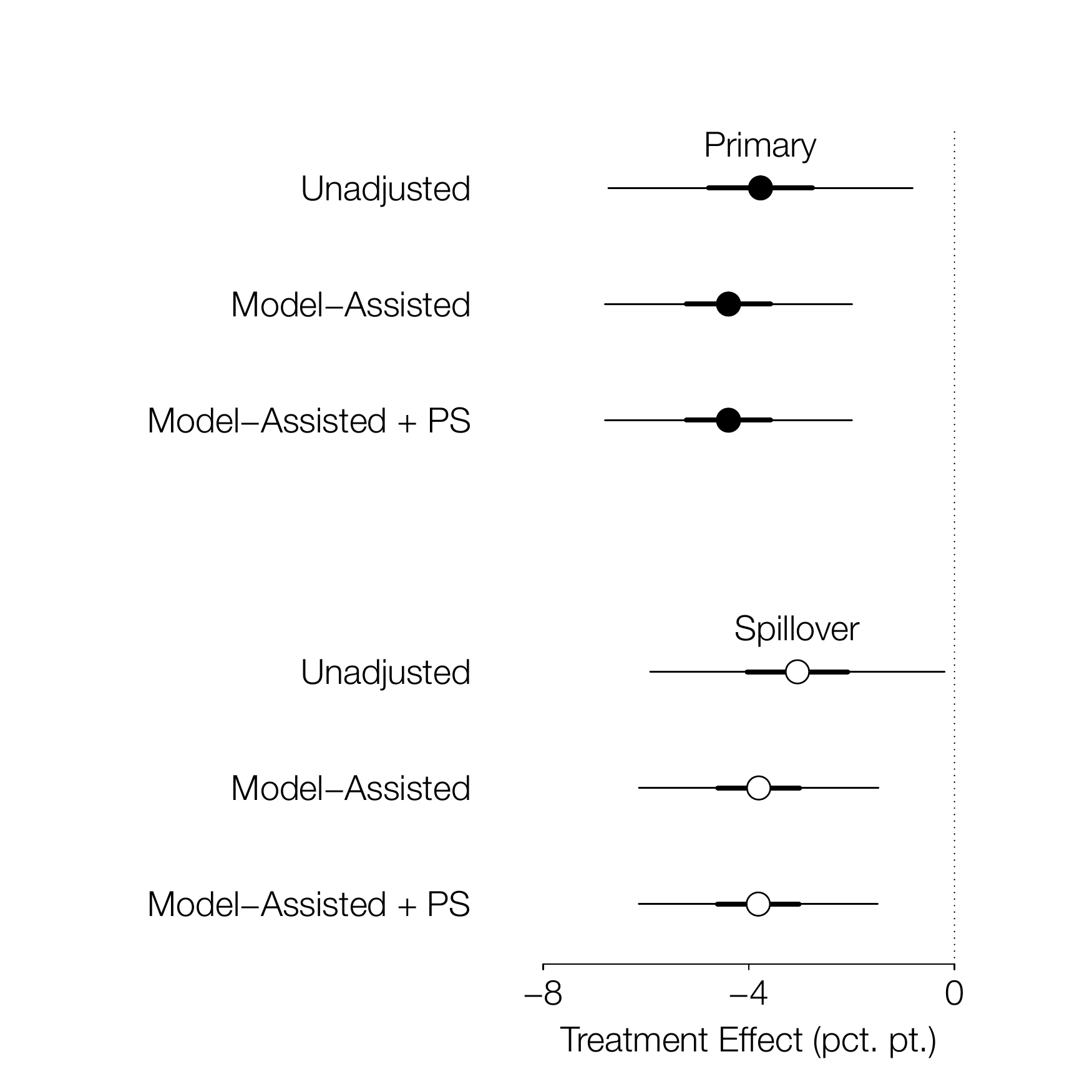}	
		\caption{Binary outcome: chronically absent}
	\end{subfigure}%
	~\begin{subfigure}[b]{0.5\textwidth}
		\centering
		\includegraphics[width = \textwidth]{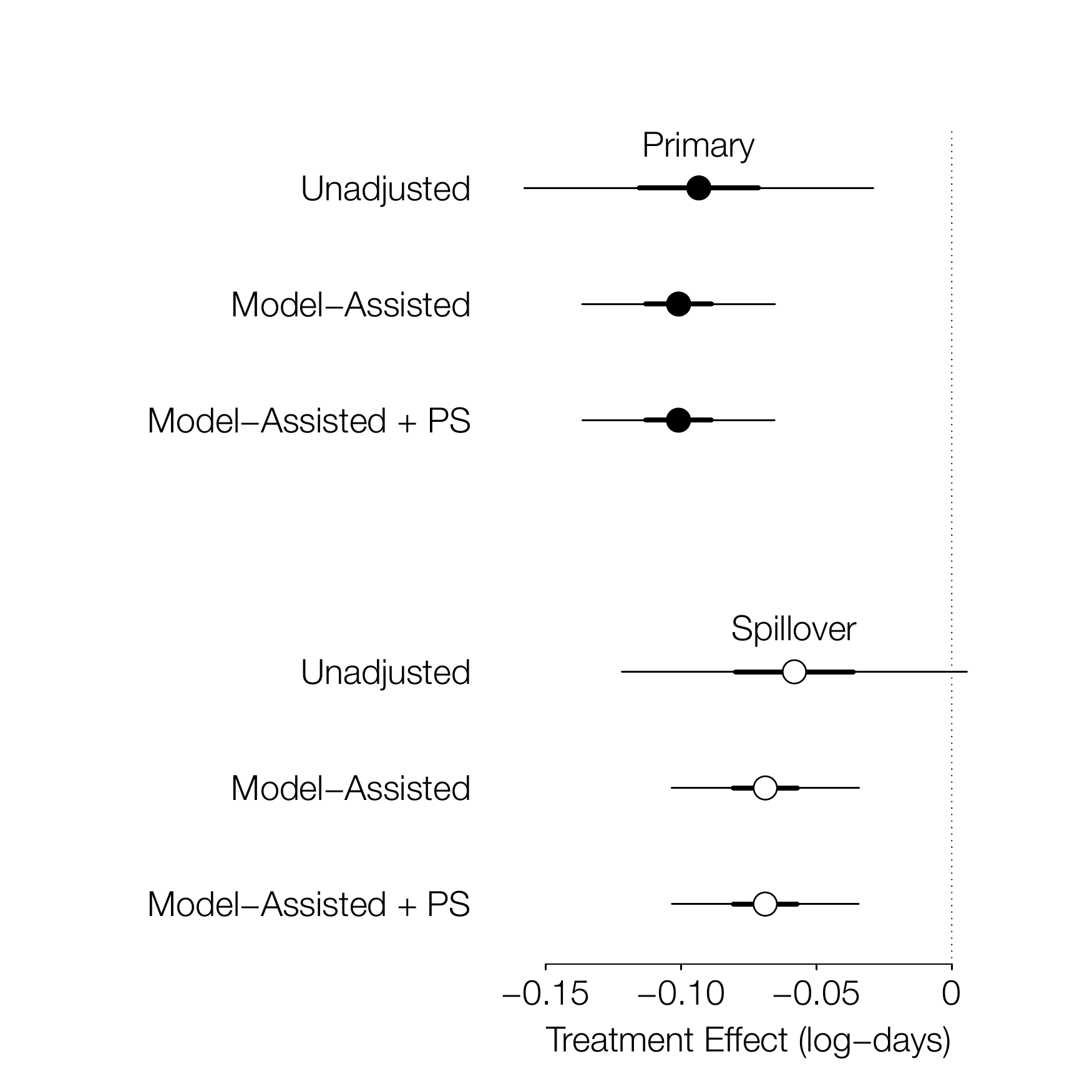}	
		\caption{Continuous outcome: log(Days + 1)}
	\end{subfigure}
	\caption{Treatment effect estimates and 95\% confidence intervals for primary (filled-in circles) and spillover (open circles) effects for individual-weighted estimands, unadjusted, with model-assisted estimation, and with post-stratification (PS) by household size.}
	\label{fig:SDP_cov_adj}	
\end{figure}

Next, Figure~\ref{fig:SDP_cov_adj} shows covariate-adjusted estimates for individual-weighted estimands. First, we take advantage of the fact that there is a natural holdout sample in the experiment as analyzed; see Section~\ref{section:data}.
To obtain $\widehat{r}_{ij}(\gamma)$, we regress the outcome on covariates listed in Table~\ref{tbl:cov_balance} as well as student grade (categorical). Results do not appear sensitive to the particular choice of model. The resulting point estimates in Figure~\ref{fig:SDP_cov_adj} are largely unchanged, if slightly larger in magnitude than the unadjusted estimates. The standard errors, however, are meaningfully smaller, especially for the continuous outcome: 0.018 log-days for the model-assisted estimator versus 0.033 log-days for the unadjusted estimator. Next, we can combine model-assisted estimation with post-stratification, though the results are essentially identical. Finally, while do not have theoretical guarantees for covariate adjustment in a two-stage experiment, classical regression adjustment is nearly identical to the model-assisted estimation with the holdout sample.

In the end, we find strong evidence of intra-household spillover for the attendance intervention. This pattern holds with and without covariate adjustment, though the covariate-adjusted estimates are more precise. This underscores that merely focusing on the primary effect significantly under-estimates the impact and cost effectiveness of the intervention.~\citet{Rogers_Feller_SDP} report costs of around \$6.60 per household. In our sample, the primary effect is around 1.2 days and the overall effect (i.e., the weighted average of the primary and spillover effects) is around 1 day. Thus, if we only consider the primary effect, the cost for each additional student day is \$6.60 / 1.2 days $\approx$ \$5.50 / day. By contrast, if we also consider spillovers, the cost for each additional student day is \$6.60 / ($2.2 \cdot 1$ day) $\approx$ \$3 / day, where $\bar{n} = n^+/N \approx 2.2$ is the average number of students per household in our sample.

\section{Discussion\label{section:discussion}}
Two-stage randomizations are increasingly common designs in settings with interactions between units. This paper addresses several practical issues in analyzing such designs. First, we address issues that arise when household sizes vary. Second, we demonstrate that regression can yield identical point- and variance-estimates to those derive from fully randomization-based methods. Methodologically, we believe that this is a useful addition to the literatures on both causal inference with interference and randomization-based inference. Substantively, we find convincing evidence of spillover effects of a large-scale attendance intervention.

There are several directions for future work. First, we are actively exploring covariate adjustment in this and other settings with more complex randomization schemes. The model-assisted approach is one such option, but many are possible~\citep{lin2013agnostic, aronow2013class}. In particular, extending the asymptotic 
results of~\citet{liu2014large} to incorporate covariates would be fruitful. Second, there is an open question of how to separately test the null hypotheses for no primary and no spillover effects in this type of design. Recent work from~\citet{athey2015exact} offers one promising direction. Third, it will be useful to extend these results to other, related designs. For example,~\citet{Weiss:2016dm} discuss an interesting setting in which random assignment occurs at the individual level but individuals are then administered treatment in groups (such as in group therapy).~\citet{kang_imbens_peer_encourage} propose a ``peer encouragement'' design, which extends the two-stage randomization considered here to consider noncompliance. Fourth, we anticipate additional connections with non-randomized studies that mimic a two-stage randomized design, such as~\citet{hong2006evaluating} and~\citet{perez2014assessing}. 

Finally, we believe one promising direction for future work is to relax the stratified interference assumption while retaining partial interference. For example,~\citet{paluck2016changing} collect detailed social network data for students within schools; the authors assume no interference between schools but leverage the within-school network structure for inference.~\citet{Arpino:2016du} offer another possibility, using covariates to construct interference patterns between units. In the end, we hope that the results we give here will lead to increased use of two-stage randomized designs in practice.

\clearpage
\singlespacing
\bibliographystyle{chicago}
\bibliography{ref}

\end{document}